\documentclass[aps,prb,twocolumn,showpacs]{revtex4-1}
\usepackage{bm}
\usepackage{amsmath}
\usepackage{graphicx}
\usepackage{natbib}

\begin{document}
\title{Anharmonic properties from a generalized
third order ab~initio approach: theory and applications to graphite and graphene}
\author{Lorenzo Paulatto}
\email[]{lorenzo.paulatto@impmc.upmc.fr}
\author{Francesco Mauri}
\author{Michele Lazzeri}
\affiliation{IMPMC, Universit\'{e} Pierre et Marie Curie, CNRS, 4 place Jussieu, F-75005 Paris, France}

\date{\today}

\begin{abstract}
We have implemented a generic method, based on the 2n+1 theorem within
density functional perturbation theory, to calculate the anharmonic
scattering coefficients among three phonons with arbitrary
wavevectors.  The method is used to study the phonon
broadening in graphite and graphene mono- and bi-layer.
The broadening of the high-energy optical branches is highly nonuniform and presents a
series of sudden steps and spikes.
At finite temperature, the two linearly dispersive acoustic branches TA and LA of graphene
have nonzero broadening for small wavevectors.
The broadening in graphite and bi-layer graphene is, overall, very similar to the
graphene one,  the most remarkable feature being the broadening of the
quasi acoustical ZO' branch.
Finally, we study the intrinsic anharmonic contribution to the thermal conductivity 
of the three systems, within the single mode relaxation time approximation. We find
the conductance to be in good agreement with experimental data for the out-of-plane direction
but to underestimate it by a factor 2 in-plane.
\end{abstract}



\maketitle

\section{Introduction}
\label{introduction}
Thermal transport is currently attracting much attention; the main applications 
of interest are materials
for thermoelectric energy conversion~\cite{mahan97,*disalvo99}
and materials used for thermal
dissipation in microelectronics~\cite{cahill03}.
While, in the first case, the goal is
to engineer the smallest possible thermal conduction, in the second
case a good thermal conduction is required.  In general, our
understanding of thermal properties is heavily based on theoretical
modeling and the use of precise and reliable approaches,
such as the \textit{ab initio} computational methods, is highly desirable.

The presence of a temperature gradient in a solid induces a heat flux. The heat
carriers can be lattice vibrations (phonons) or electronic
excitations.  In general, lattice conduction is the dominant mechanism
in the presence of an electronic gap (semiconductors and insulators)
or when the gap is zero but the density of electronic states at the
Fermi level is small (semi-metals). Lattice thermal resistance is then
dictated by phonon scattering, which can be induced by
extrinsic mechanisms (isotopic disorder, structural defects,
finite-size of the crystals, etc.) or by intrinsic ones (anharmonic phonon-phonon
scattering). Determining the intrinsic anharmonic scattering
is in itself a very complex task; it has been attempted \textit{ab
 initio}, within density functional theory (DFT), using finite difference derivation \cite{shobhana,tang10,*tang11} or molecular dynamics techniques \cite{donadio} or from linear response theory.\cite{broido07,ward09,*ward10,garg11,bonini12}

In a crystal, the intrinsic lattice thermal conduction can be obtained by knowing
harmonic phonon energies and anharmonic phonon-phonon scattering coefficients.
Harmonic phonon energies are determined by the second order derivative of
the system total energy, with respect to atomic displacements.
This second derivative can be efficiently calculated \textit{ab initio} by using density
functional perturbation theory (DFPT)~\cite{dfpt}, which
allows the determination of the phonon dynamical matrix 
for an arbitrary {\bf q} wavevector in the Brillouin zone.
DFPT is implemented in the {\sc Quantum ESPRESSO} package~\cite{wwwqe,*qe},
within the plane-waves and pseudopotential approaches.
The anharmonic scattering coefficients can be determined by the third
order derivative of the energy with respect to three phonon
perturbations, characterized by the wavevectors {\bf q}, ${\bf q}'$, ${\bf q}''$.
For the thermal transport problem, it is necessary to know these
derivatives with respect to three arbitrary wavevectors 
(possibly ${\bf q}\ne{\bf 0}$, ${\bf q'}\ne{\bf 0}$, ${\bf q''}\ne{\bf 0}$),
with the only condition ${\bf q}+{\bf q'}+{\bf q''}={\bf G}$,
where {\bf G} is a reciprocal lattice vector.
In principle, these coefficients can be obtained within DFPT by using
the, so called, ``2n+1'' theorem as formulated by Ref.~\onlinecite{gonzeron}. This theorem allows us to access the 3rd derivative of the total energy by using only the 1st derivative of ground state density and wavefunctions; contrary to the finite differences approach, we do not have to perform expensive supercell calculations.

The first implementations of the ``2n+1'' approach were limited to
the scattering of one zero-momentum phonon towards two phonons
with opposite arbitrary momenta ({\bf 0}, {\bf -q}, {\bf q}).
Ref.~\onlinecite{debernardi94}
implemented this approach for insulating and semiconducting materials.
Later, Ref.~\onlinecite{lazzeri02} generalized the approach to metals and zero-gap materials.
The ({\bf 0}, {\bf -q}, {\bf q}) anharmonic coefficients can be computed
within {\sc Quantum ESPRESSO}~\cite{wwwqe,*qe}, using the 
{\sc d3} code which was developed in Ref.~\onlinecite{lazzeri02}.
These coefficients can be easily used to compute the anharmonic broadening
of a ${\bf q=0}$ phonon (see \textit{e.g.} Ref.~\onlinecite{lazzeri03}).
By using a super-cell approach one can also compute the broadening of
phonons having {\bf q} commensurate with the super cell.
However, the super-cell approach (which was used
in Ref.~\onlinecite{garg11,bonini12,lazzeri03}) can be computationally very demanding.
Recently, Ref.~\onlinecite{deinzer03} further extend the method to three
arbitrary phonons, (${\bf q}, {\bf q'}, {\bf q''}$), although only for
insulators/semiconductors. This was done within a proprietary non-publicly-available software.

We have developed and implemented an extension of
the {\sc d3} code of the {\sc Quantum ESPRESSO} software
to compute, within the DFPT ``2n+1'' approach,
the three-phonons anharmonic coefficients
for three arbitrary wavevectors $({\bf q}, {\bf q'}, {\bf q''})$
for  insulators/semiconductors and also for metallic or zero gap
systems.
The coefficients thus obtained can be used in a straightforward way to
compute the anharmonic broadening of a phonon with an arbitrary
wavevector {\bf q} and the intrinsic thermal conductivity within
the single-mode relaxation time approximation (SMA).
~\cite{srivastava74,*asenpalmer97,*cao04,*khitun01}
The first applications of the method are devoted to 
graphite, graphene and graphene bi-layer.
Indeed, the thermal properties of these systems have attracted
significant attention,~\cite{balandin,seol10} being $sp^2$ carbon
systems excellent thermal conductors.

Sec.~\ref{sec2} describes the method.
Sec.~\ref{sec3} reports the results and the discussion.
Conclusions are summarized in Sec.~\ref{sec4}.

\section {Method}
\label{sec2}
In Sec.~\ref{sec_anh} we provide the expressions for the phonon anharmonic
broadening and for the phonon thermal conduction.
In Sec.~\ref{sec_compdet} the method is briefly described
and we provide the relevant computational details.
A more detailed description of the method is reported in the Appendix.

\subsection{Anharmonic decay and thermal transport\label{sec_anh}}
Let us consider the total energy for a crystal 
${\cal E}^{\rm tot}(\{v_{{\bf R},s,\alpha}\})$, where
$v_{{\bf R},s,\alpha}$ is
the displacement from the equilibrium position of the
$s$ atom along the $\alpha$ Cartesian coordinate in the
unit cell identified by the lattice vector ${\bf R}$.
We define
\begin{equation}
u_{{\bf q},s,\alpha} = 
\frac{1}{N}\sum_{\bf R} e^{-i{\bf q}\cdot{\bf R}}
v_{{\bf R},s,\alpha},
\label{equq}
\end{equation}
where the sum is performed on the lattice vectors $\{{\bf R}\}$
and $N$ is the number of cells involved in the summation.
We define the dynamical matrix
\begin{equation}
D_2\left(
\begin{smallmatrix}
&{\bf q}\\
s&s'\\
\alpha&\alpha'
\end{smallmatrix}
\right)=
\frac{1}{N}
\frac{\partial^2 {\cal E}^{\rm tot}}{
\partial u_{{\bf -q},s,\alpha}
\partial u_{{\bf q},s',\alpha'}
},
\label{eqdyn}
\end{equation}
the angular frequency $\omega_{{\bf q},j}$ of a phonon with wavevector
{\bf q} and branch index $j$ is obtained by solving
\begin{equation}
\sum_{s',\alpha'}
\frac{1}{\sqrt{m_sm_{s'}}}
D_2\left(
\begin{smallmatrix}
&{\bf q}\\
s&s'\\
\alpha&\alpha'
\end{smallmatrix}
\right)
z_{s',\alpha'}^{{\bf q},j}=
\omega^2_{{\bf q},j}
z_{s,\alpha}^{{\bf q},j},
\end{equation}
where $z$ are the orthogonal phonon eigenmodes
normalized in the unit cell and $m_s$ is the atom mass.
We define the three-phonon scattering coefficients as
\begin{equation}
V^{(3)}_{{\bf q}j,{\bf q'}j',{\bf q''}j''}=
\frac{1}{N}
\frac{\partial^3{\cal E}^{\rm tot}}
{
\partial X_{{\bf q},j}
\partial X_{{\bf q'},j'}
\partial X_{{\bf q''},j''}
},
\label{eqv3}
\end{equation}
where
\begin{equation}
\frac{\partial}{\partial X_{{\bf q},j}}=
\sum_{s,\alpha}
\sqrt{\frac{\hbar}{2m_s\omega_{{\bf q},j}}}
z_{s,\alpha}^{{\bf q},j}
\frac{\partial}{\partial u_{{\bf q},s,\alpha}}.
\label{eq5}
\end{equation}
$V^{(3)}$ has the dimension of an energy and does not depend on $N$, while $X_{{\bf q},j}$
is adimensional.
Because of the translational symmetry of the crystal, the coefficients
$V^{(3)}$ from Eq.~\ref{eqv3} are $\ne 0$ only when
${\bf q}+{\bf q'}+{\bf q''}={\bf G}$, where {\bf G} is any
reciprocal lattice vector.

With these definitions, the lifetime
due to anharmonic phonon--phonon interaction, $\tau_{{\bf q}j}$,
and the corresponding broadening $\gamma_{{\bf q}j}$ (full width at half maximum)
of the phonon $({\bf q}j)$ are~\cite{lazzeri03}:
\begin{equation}
\begin{split}
\frac{1}{\tau_{{\bf q}j}(T)}={}&\gamma_{{\bf q}j}(T)\\
={}&\frac{\pi}{\hbar^2 N_q}
\sum_{{\bf q'},j',j''}
\left|V^{(3)}_{{\bf q}j,{\bf q'}j',{\bf
      q''}j''}\right|^2 \\
&\times
\Big[
(1+n_{{\bf q'}j'}+n_{{\bf q''}j''}))
\delta(\omega_{{\bf q}j}-\omega_{{\bf q'}j'}-\omega_{{\bf
    q''}j''}) \\
&\phantom{\times\Big[}+2(n_{{\bf q'}j'}-n_{{\bf q''}j''})
\delta(\omega_{{\bf q}j}+\omega_{{\bf q'}j'}-\omega_{{\bf q''}j''})
\Big].
\label{eq_tauanh}
\end{split}
\end{equation}
Where $T$ is the temperature,
$n_{{\bf q}j}$ is the Bose-Einstein statistics occupation of 
phonon $({\bf q}j)$, and $\delta(x)$ is the Dirac distribution.
The sum is performed over a sufficiently fine grid of $N_q$ {\bf q}-points
in the Brillouin zone (BZ)  and ${\bf q''}=-{\bf q}-{\bf q'}$.
$\tau_{{\bf q}j}$ and $\gamma_{{\bf q}j}$
depend on $T$ only through the
phonon occupations $n$.

The r.h.s. of Eq.~\ref{eq_tauanh} is usually interpreted as the sum of
scattering processes in which a phonon of wavevector {\bf q} decays into two
phonons $-{\bf q}'$, $-{\bf q}''$, (third line of Eq.~\ref{eq_tauanh})
or in which the phonon {\bf q} coalesces with $-{\bf q}'$ and emits
$-{\bf q}''$ (fourth line of Eq.~\ref{eq_tauanh}).
The energy conservation of the processes are guaranteed by the
Dirac delta.
One can also distinguish between Normal and Umklapp processes:
by choosing {\bf q} and $-{\bf q}'$ such that they belong to the first BZ,
the scattering is Normal when also ${\bf q''}=-{\bf q}-{\bf q'}$
belongs to the first BZ; on the contrary, when ${\bf q''}$ does not
belong to the first BZ, the scattering is Umklapp.

By knowing the anharmonic scattering coefficients, Eq.~\ref{eqv3}, one can
determine the lattice thermal conductivity within the framework of
the Boltzmann transport equation (BTE) for phonons~\cite{ziman}.
In general, an exact solution of the BTE is a 
difficult task; a commonly used approximation to the problem is
the so-called single mode relaxation time
approximation~(SMA)\cite{garg11,srivastava74,asenpalmer97,cao04,khitun01}.
Within the SMA, the lattice thermal conductivity tensor becomes:
\begin{equation}
\kappa^{\alpha,\beta}_{\rm L}=\frac{\hbar^2}{N_q\Omega K_{\rm B} T^2}
\sum_{{\bf q}j}
c^\alpha_{{\bf q}j}
c^\beta_{{\bf q}j}
\omega^2_{{\bf q}j}
n_{{\bf q}j}
(n_{{\bf q}j}+1)
\tau_{{\bf q}j}.
\label{eq_k}
\end{equation}
Here, $\Omega$ is the volume of the unit cell, $K_{\rm B}$ is the Boltzmann
constant and $c^\alpha_{{\bf q}j}$ is the phonon group velocity of mode
$({\bf q}j)$ along Cartesian direction $\alpha$:
$c^\alpha_{{\bf q}j}=d\omega_{{\bf q}j}/(dq_\alpha)$.
The SMA conductivity from Eq.~\ref{eq_k} can be obtained in a
straightforward way once the anharmonic lifetimes $\tau_{{\bf q}j}$
have been computed from Eq.~\ref{eq_tauanh}.
$\kappa^{\alpha,\beta}_{\rm L}$ is a $3\times3$ tensor which takes into
account the possible anisotropies and transversal
conductance. However, in high-symmetry crystals, as graphene and graphite,
the off-diagonal elements are zero, if two axes lie in the graphene plane.
Moreover, in both graphene and graphite the two in-plane $xx$ and $yy$
components are identical. The out-of-plane $zz$ component is not 
defined in the bidimensional graphene systems, but it is meaningful in
graphite.

The validity limits of the SMA are discussed in literature~\cite{garg11,srivastava74,asenpalmer97,cao04,khitun01}.
Here, we wish to remind that, for a generic material, the SMA is 
expected to be valid  (that is, to provide the correct solution to the BTE)
at room conditions and to break down only at very small temperatures
(see, \textit{e.g.} Ref.~\onlinecite{ward09,ward10}).

\subsection{Density-functional theory calculation}
\label{sec_compdet}

Calculations of the phonon properties are done within density
functional perturbation theory~\cite{dfpt} as implemented
in Ref.~\onlinecite{qe}.  The third order coefficients defined in
Eq.~\ref{eqv3} are computed using a code which has been developed for
the present work. This code has been written on the top of a previous
less general implementation available within
the~{\sc Quantum ESPRESSO}
package: the {\sc d3} code, which was implemented in
Ref.~\onlinecite{lazzeri02}.  
The method is described in detail in
Appendix~\ref{app1}, and ~\ref{app_implementation}.

\begin{figure}
\resizebox{0.85 \columnwidth}{!}
{\includegraphics{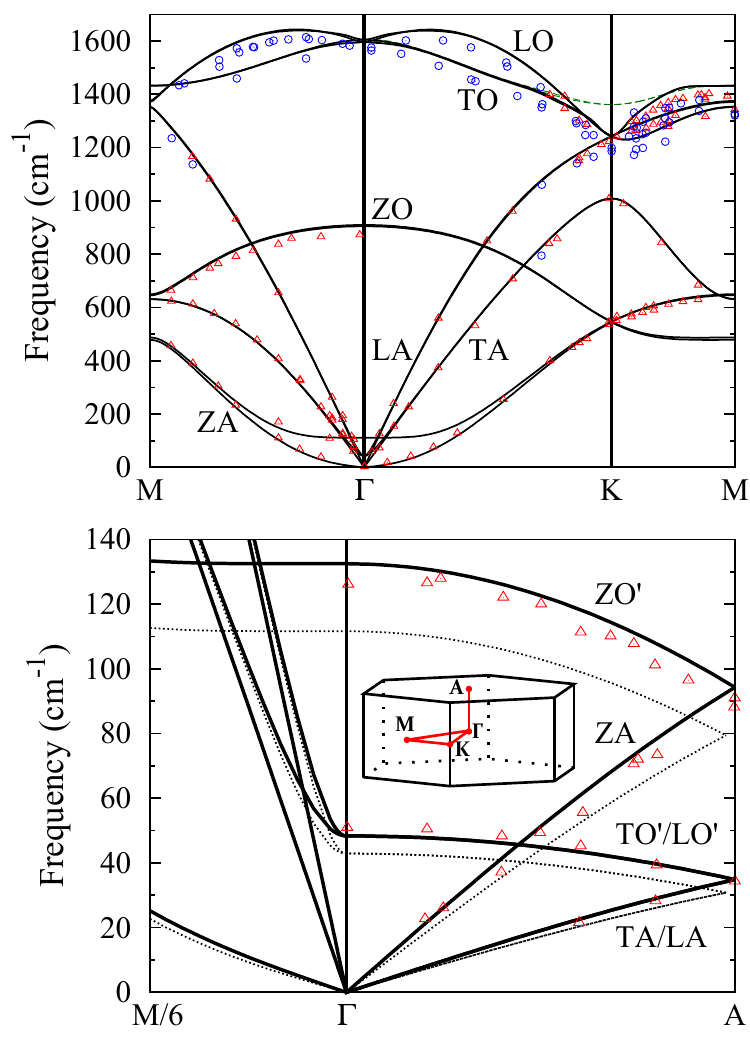}}
\caption{\label{fig0}
(Color online) Graphite phonon dispersion.
Symbols are measurements from Refs.~\onlinecite{maultzsch04,mohr07}.
Lines are calculations.
In the upper panel, the TO optical branches are plotted two times.
The solid (black) lines include GW corrections of the electron-phonon interaction.
On the contrary the dashed (green) lines are done using standard DFT and they
are shown only for comparison.
Lower panel: Solid lines are done by using $c/a=2.664$;
dashed lines are done with $c/a=2.727$ and they are shown only for comparison.
In the inset we recall the high symmetry points naming convention.
In both panels, the solid lines correspond to the calculations
used throughout the paper.}
\end{figure}

We use local-density approximation and
the carbon atom is described by a norm-conserving pseudopotential
which includes four electrons in valence. Plane waves kinetic energy
cutoff is $90$~Ry.  For all the systems, the in-plane lattice
parameter is $a=2.44$~\AA, which is the theoretical equilibrium value
for graphite.  For graphite, we use $c/a=2.664$. This value, which is only slightly different from
the experimental value $c/a=2.727$, is chosen phenomenological
to accurately reproduce the low frequency phonon dispersion along the
${\bm \Gamma}$-{\bf A} direction (see the discussion below).
For graphene, the periodic replicas of the
planes are spaced along the $z$ direction with $7$~\AA~of vacuum.  The
two layers of the graphene bilayer are spaced with the inter-planar
distance of bulk graphite; periodic images are then spaced with
$7$~\AA~of vacuum.

The computational parameters are listed in Appendix~\ref{comp_det}.
We remind here that the electronic integration has to be done
with a small value of smearing (and a consequent
fine-grained {\bf k}-point grid) due to the presence of
a Kohn anomaly for the highest optical branch near {\bf K}~\cite{piscanec}
(usually called TO).
The phonon frequencies $\omega_{{\bf q}j}$ and the third-order coefficients $V^{(3)}$, used in
Eqs.~\ref{eq_tauanh} and~\ref{eq_k}, are calculated in a slightly
different way.
On one hand, phonon energies are corrected using an \textit{ad hoc} procedure
(based on DFT+GW renormalization of the electron-phonon interaction as in 
Ref.~\onlinecite{lazzeri08}, see Appendix~\ref{comp_det}).
This correction affects only the TO branch, it does not touch the other branches,
and it provides better agreement with measurements, Fig.~\ref{fig0}.
On the other hand, the third-order coefficients are computed within
standard (less precise) DFT.
The using of these two different procedures for the
$\omega_{{\bf q}j}$ and $V^{(3)}$ calculations is not consistent. However,
this should not affect the results in a major way:
phonon broadening results from
a sum over different processes which are selected by energy conservation
enforced by the two Dirac $\delta$ in Eq.~\ref{eq_tauanh}. The intensity of the
processes is then proportional to the square of the $V^{(3)}$ coefficients.
Consequently, the computational accuracy of $\omega_{{\bf q}j}$ and that of
the $V^{(3)}$ coefficients affect the result in a very different way.
An error in the phonon dispersion can affect the lifetime in a not predictable way and, thus,
a special care should be taken into finding the best possible description
of the phonon dispersion. The same care is not strictly necessary for the third order calculations.

Fig.~\ref{fig0} compares measured with calculated phonon dispersions for 
graphite. Notice that plain DFT calculations do not provide a satisfactory description
of the highest optical TO branch near {\bf K}, while DFT+GW ones do much better.
The lower panel of Fig.~\ref{fig0} shows in detail the low frequency dispersion.
This region is characterized by the splitting of the acoustic phonon branches of 
the two graphene planes in the graphite unit cell.
These branches are particularly sensitive to the actual value of $c/a$.
In particular, in that region, by changing the lattice parameters from $c/a=2.664$ (which is the value used
throughout the paper) to $c/a=2.727$ the value of the phonon branches change by almost 14\%.

Actual DFT calculations are done on a relatively coarse grid of {\bf
  q} wavevectors, described in Appendix~\ref{comp_det}.
The dynamical matrices and the third order coefficients, necessary to compute 
the broadening and the thermal conductivity
(Eqs.~\ref{eq_tauanh} and ~\ref{eq_k}), are then obtained 
on a finer grid via the Fourier interpolation technique described in Appendix~\ref{app2}.
Eqs.~\ref{eq_tauanh} and ~\ref{eq_k} are evaluated by performing the sum over a discrete
grid of ${\bf q}$ points and by substituting the $\delta(x)$ 
with a Gaussian function characterized by an artificial smearing $\chi$.
This approximation is valid as long as $\chi$ is smaller than the thermodynamic
fluctuation, which is of order $K_{\rm B} T$.
The grids and the $\chi$ values are specified in Appendix~\ref{comp_det}. 
Here we just remark that the results shown in Sects.~\ref{sec31} and ~\ref{sec32}
are obtained using a particularly fine-grained sampling.
This is only necessary to produce the very
sharp features which are present in the broadening of the higher optical bands or to
produce the correct behavior of the broadening of the acoustic branches in the vicinity
of ${\bm \Gamma}$.
Indeed, a much coarser grid is sufficient for most applications, included those described in
Sec.~\ref{sec33}.

\section{Results and discussion}
\label{sec3}
This section reports and describes the results.
Section~\ref{sec31} analyzes the anharmonic phonon broadening in the
graphene monolayer. Special relevance is given to the three
acoustic branches which are the most important for the thermal transport.
Sec.~\ref{sec32} analyzes the broadening in graphene bilayer and graphite.
Sec.~\ref{sec33} is dedicated to thermal transport.

\subsection{Graphene phonon broadening}
\label{sec31}
\begin{figure*}[t!]
\resizebox{2.0 \columnwidth}{!}
 {\includegraphics{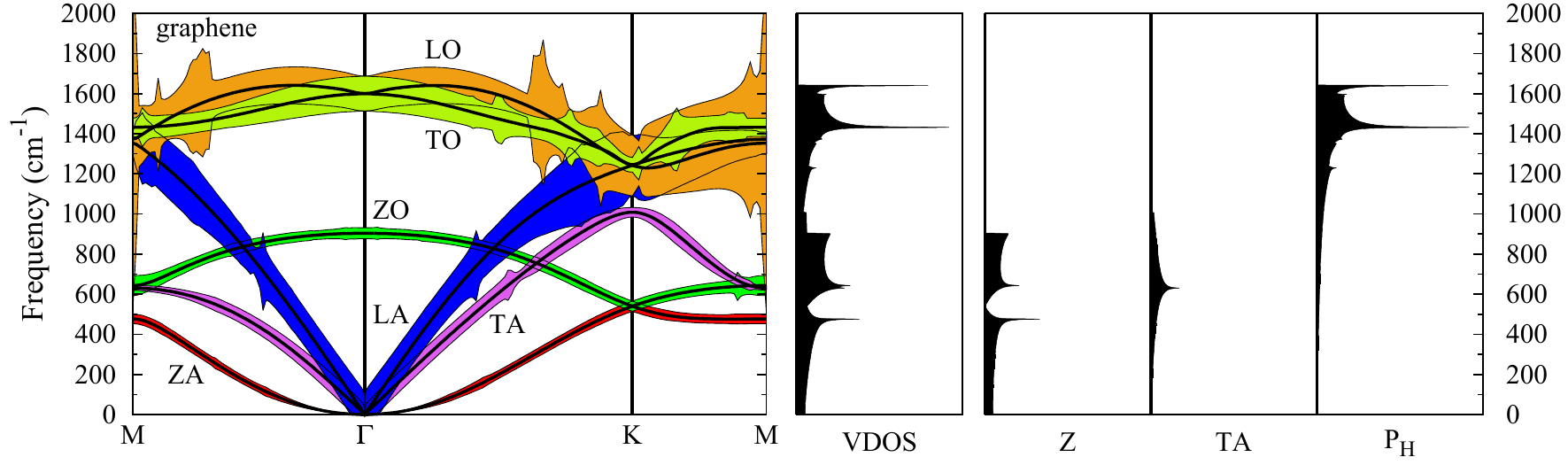}}
\caption{\label{gene} 
(Color online) Calculated graphene phonon dispersion. Each phonon branch is
represented with a variable-width filled band:
the graphical width is equal to the respective anharmonic
broadening at $300$~K, expressed in cm$^{-1}$ and magnified by a
factor 100.
The vibrational density of states (VDOS) is also shown, together with
its decomposition over groups of disentangled branches
labeled as Z (corresponding to the ZA and ZO branches), 
TA and ${\rm P_H}$ (corresponding to LA, TO, and LO).}
\end{figure*}
\begin{figure}[b!]
\resizebox{1.00 \columnwidth}{!}
{\includegraphics{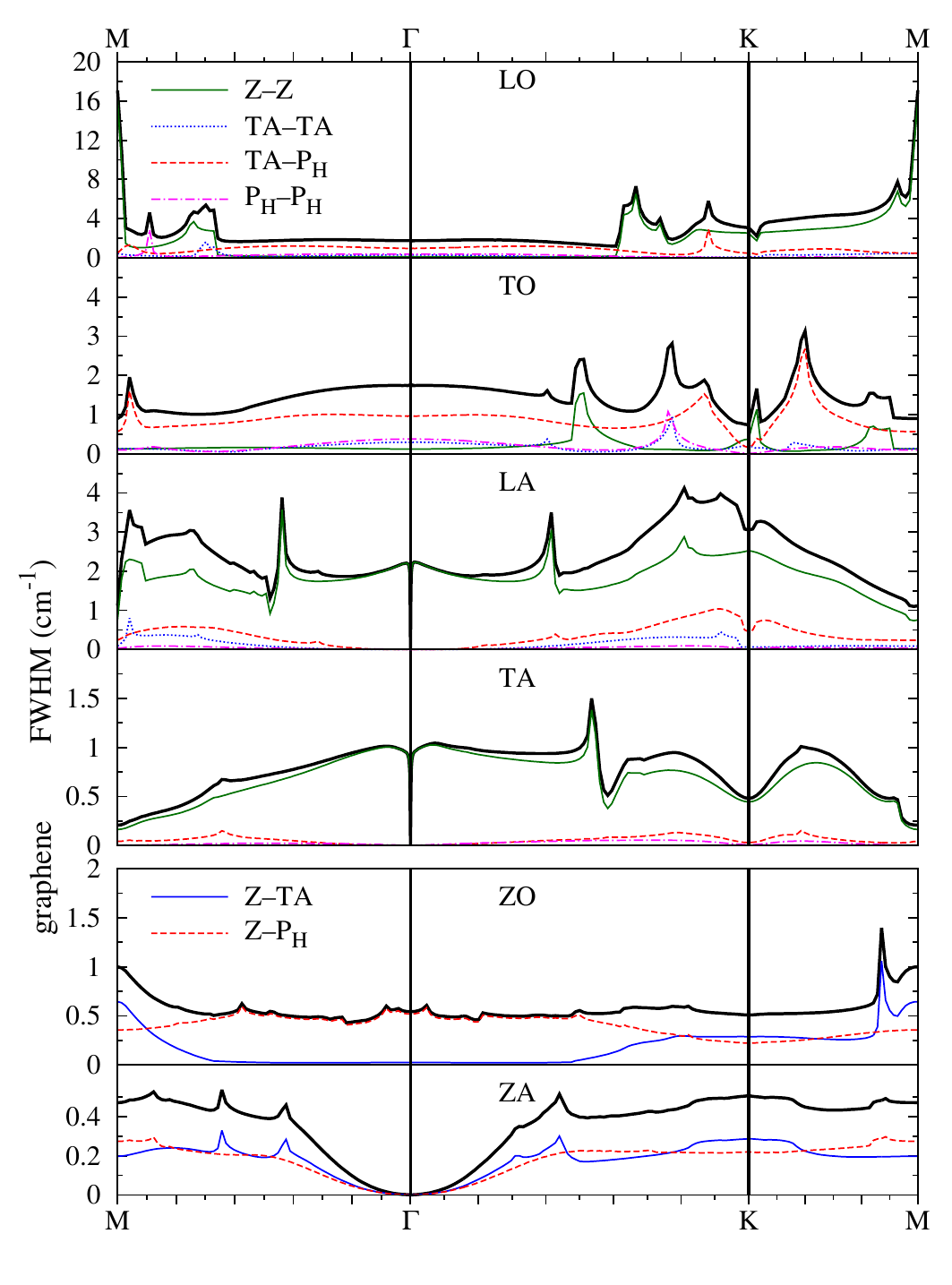}}
\caption{\label{gene2}
(Color online)
Graphene anharmonic phonon broadening (FWHM) at 300~K,
for each phonon branch
(labeled as in Fig.~\ref{gene}), along high symmetry lines.
The total broadening (solid thick line) is decomposed depending on the
character of the final states which are labeled as Z, TA, and  ${\rm P_H}$
(see the text): \textit{e.g.} TA-${\rm P_H}$ corresponds to a decay
involving one TA and one ${\rm P_H}$ phonon.}
\end{figure}

Fig.~\ref{gene} shows the calculated graphene phonon dispersion,
the respective anharmonic broadening and the vibrational density of
states (VDOS).
The branches are labeled in the usual way.\cite{mounet}
There are three acoustic branches (ZA, TA, LA) and three optical branches
(ZO, TO, LO).
ZA and ZO correspond to an atomic motion perpendicular to the 
graphene plane ($z$ direction), all the other branches are polarized parallel to the plane.
In the vicinity of ${\bm \Gamma}$, TA and TO are quasi transverse, while LA and LO
are quasi longitudinal.
In the following, these labels will be used to classify the branches all along the high
symmetry lines (as in Fig.~\ref{gene}), although this distinction is
not meaningful for an arbitrary wavevector in the Brillouin
zone (BZ).
Because of symmetry, the modes perpendicular polarized (ZA and ZO) are separated from the
others all over the BZ. Moreover, the TA branch is always well separated from the other parallel
polarized branches (labeled as ${\rm P_H}$).
In Fig.~\ref{gene}, we can, thus, separate the VDOS in three distinct components labeled as Z, TA, and ${\rm P_H}$.
The two dimensional character of the phonon dispersion is associated
with some specific features.
The ZA branch is quadratic near ${\bm \Gamma}$ and, thus, in the limit $\omega\rightarrow0$
the VDOS does not go to zero (Fig.~\ref{gene}).
The presence of a local maximum in the phonon dispersion
(as the one at 1008 cm$^{-1}$ for the TA branch near \textbf{K} or the one at 904 cm$^{-1}$ for
the ZO one near ${\bm \Gamma}$)
is associated with a step in the VDOS.
The presence of a saddle point in the dispersion 
(as those at 477~cm$^{-1}$, 631~cm$^{-1}$, 643~cm$^{-1}$, and 1432~cm$^{-1}$ at the \textbf{M} point)
is associated with a sharp peak in the VDOS.

Fig.~\ref{gene2} reports in more detail the calculated anharmonic
phonon broadening, along high symmetry lines, and its decomposition
into the different allowed decay channels.
For symmetry reasons, the $z$-polarized branches can only decay toward one Z and one non-Z phonons.\cite{bonini07,mingo10}
The other bands can only decay towards two phonons which are either both or neither
$z$-polarized.
The two most striking features in Figs.~\ref{gene},~\ref{gene2} are the
small $q$ behavior of the acoustic branches and the highly non uniform behavior
of the broadening.

\begin{figure}[b!]
\resizebox{1.00 \columnwidth}{!}
{\includegraphics{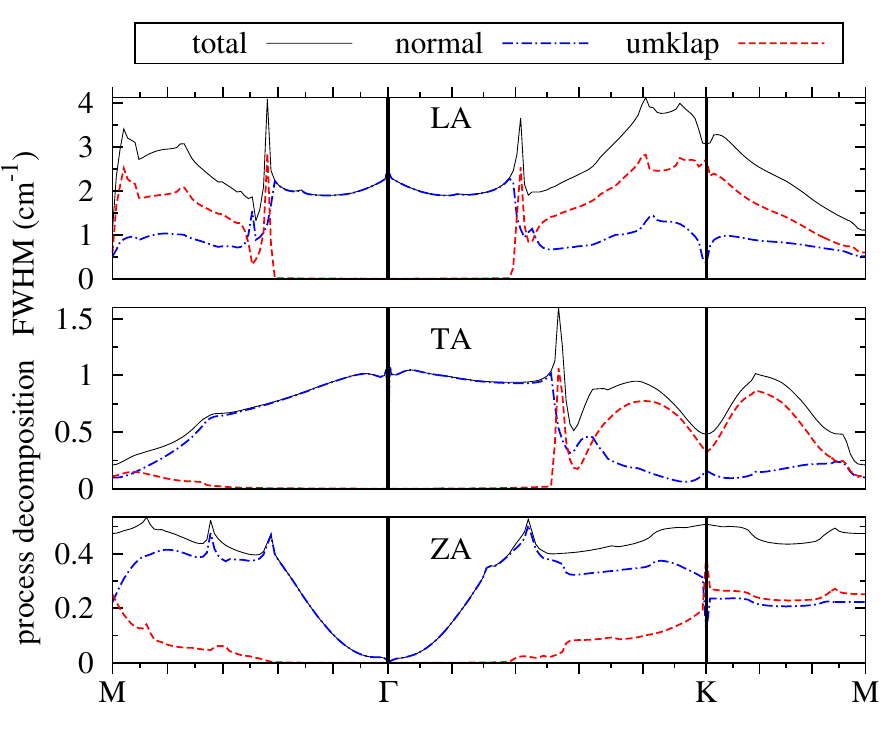}}
\caption{\label{umklapp}
(Color online) Graphene anharmonic phonon broadening for the
three acoustic branches at 300~K (same data as in Fig.~\ref{gene2}):
the broadening is decomposed in two components which are due, respectively,
to Normal and Umklapp processes.
}
\end{figure}

First, we remind that
in a three dimensional isotropic crystal, all the three acoustic
branches are linearly dispersive and one expects to observe for $q\rightarrow0$ 
a vanishing broadening.
On the contrary, at finite temperature, both TA and LA branches of the
two dimensional graphene have a nonzero broadening in the $q\rightarrow0$ limit.
This behavior is due to a decay process in which a TA (or LA) phonon decays into
two phonons both in the ZA branch, Fig.~\ref{gene2}.
This decay is entirely due to Normal scattering.
This can be seen in Fig.~\ref{umklapp}, where the broadening of the acoustic branches
is decomposed into the two components which are due, respectively to Normal and 
Umklapp processes, as defined in Sec.~\ref{sec_anh}.
Actually, one can demonstrate\cite{bonini12} that, in general, when a linearly dispersive phonon
decays into two quadratically dispersive phonons, the broadening is non
vanishing in the $q\rightarrow0$ limit
because of energy and momentum conservation.
Moreover, the quadratically dispersive ZA branch has a broadening which
is itself quadratic in $q$ around ${\bm \Gamma}$.
The ZA broadening is due to a Normal decay in which the ZA phonon decays into
one ZA phonon and one linearly dispersive, TA or LA, phonon.
Again, one can demonstrate that, in general, when a quadratically
dispersive phonon decays into one quadratically and one linearly
dispersive phonons, the broadening vanishes quadratically in the $q\rightarrow0$ limit.
We remark that, here, the anharmonic broadening has been computed
by summing over an extremely fine reciprocal-space grid (see Appendix~\ref{comp_det}).
This is necessary in order to reproduce correctly
the anomalous behavior of the LA and TA broadening for small $q$.
Far from ${\bm \Gamma}$, the details of the broadening can be correctly
reproduced by using a much coarser grid and a larger $\chi$.

\begin{figure*}[t!]
\resizebox{2.00 \columnwidth}{!}
 {\includegraphics{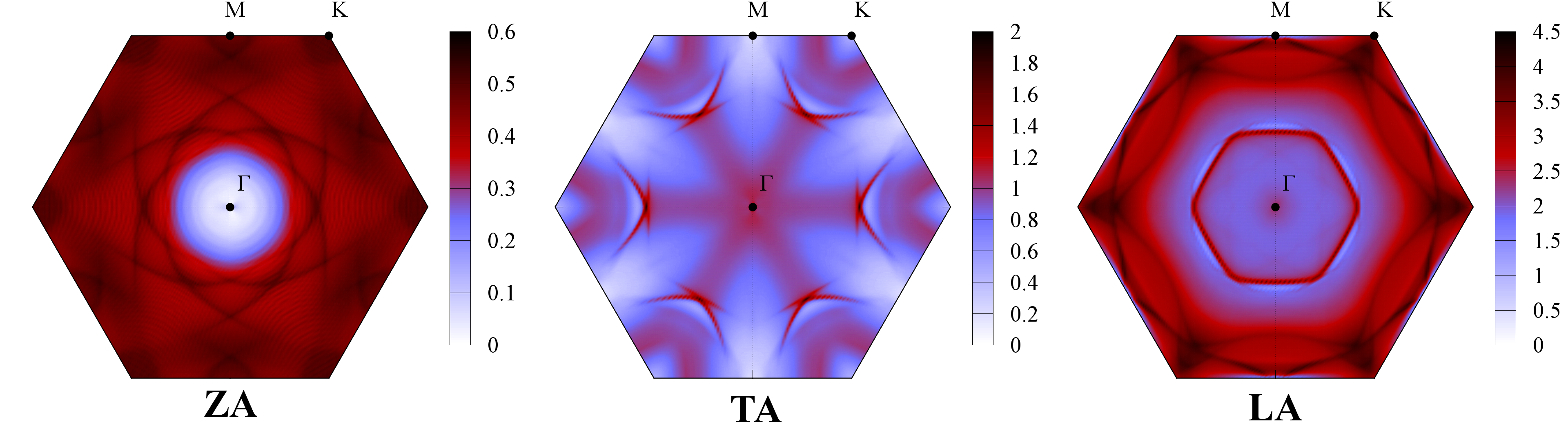}}
\caption{\label{gene-2d} (Color online) 
Graphene anharmonic phonon broadening (FWHM, in cm$^{-1}$) at 300~K,
for the three acoustic branches, over all the Brillouin zone.}
\end{figure*}

The existence of a finite broadening at small $q$ for the TA and LA acoustic
branches is problematic. Indeed, the concept itself of
phonon is meaningful only when $\omega/\gamma>1$, being $\gamma$
the broadening (\textit{i.e.} the inverse of the phonon lifetime).
From the present calculations, the condition $\omega/\gamma>1$ is satisfied
for both the TA and LA branches for $q>\overline{q}$, with
$\overline{q}=0.5\times 10^{-4} 2\pi/a_0$, being $a_0$ the in-plane lattice spacing.
Thus, for $q<\overline{q}$, the TA or LA frequency
can become smaller that the broadening. In this region the present
treatment is, obviously, not valid (see the discussion in Ref.~\onlinecite{bonini12})
and a proper treatment of the phenomenon is beyond the present scope.
In practice, however, $q<\overline{q}$ represents a tiny portion of
the Brillouin zone (the corresponding region in
Figs.~\ref{gene},~\ref{gene2} has width of the order of the thickness
of the vertical line passing through ${\bm \Gamma}$).
As a consequence, we can assume that the properties 
obtained as a sum over the Brillouin zone (such as
the thermal conductivity of Eq.~\ref{eq_k}) are not affected by a
major error.

Concerning the global appearance of Figs.~\ref{gene},~\ref{gene2},
the many sharp peaks in the broadening can be ascribed to different
mechanisms. 
Those in the high energy part of the spectrum are, in general, associated with peaks in the VDOS:
when one or both of the final states (\textit{i.e.} of the states
that meet the energy and momentum conservation requirements in Eq.~\ref{eq_tauanh}) produce a
peak in the VDOS, there the broadening will typically exhibit a peak.
For example, the large scattering probability, predicted for the
{\bf M}
point on the LO branch ($1373$ cm$^{-1}$), corresponds to a decay toward
a ZO phonon close to ${\bm \Gamma}$ ($904$ cm$^{-1}$) and a ZA phonon close
to {\bf M} ($477$ cm$^{-1}$). As the VDOS, Fig. \ref{gene},
has maximum in both region, this transition is particularly favored.
On the other hand, for {\bf q}$\gtrsim$0.66{\bf M} (along the ${\bm \Gamma}${\bf M}
direction) or for {\bf q}$\gtrsim$0.62{\bf K} (along ${\bm \Gamma}${\bf K}),
the LO broadening displays a sudden increase.
This is because, for these wavevectors, the LO energy becomes
small enough to activate the decay channel towards the ZO branch
(At {\bf q}$\sim$0.66{\bf M} and {\bf q}$\sim$0.62{\bf K} the LO phonon
decays into a ZO with the same wavevector and a ZO with 
{\bf q}$\sim{\bm \Gamma}$).

We remark that the presence of sharp peaks which are essentially
determined by energy and momentum conservation in the decay process
implies that even a small change in the phonon dispersion
used in the calculation could induce significant differences in the
calculated broadening.

Concerning the three acoustic branches
the broadening near ${\bm \Gamma}$ is almost entirely due to Normal scattering, Fig.~\ref{umklapp}.
The peaks which are observed at
{\bf q}$\gtrsim$0.44~{\bf M} and {\bf q}$\gtrsim$0.41~{\bf K}
for the LA branch,
and at {\bf q}$\gtrsim$0.65~{\bf M} and {\bf q}$\gtrsim$0.54~{\bf K} for the
TA one, are associated with the activation of Umklapp scattering
towards the ZA phonons.
To have a more comprehensive view, Fig. \ref{gene-2d} reports the 
broadening in the entire Brillouin zone.
The TA and LA branches exhibit a feature-rich behavior in a wide
region, far from ${\bm \Gamma}$, which roughly starts
at about halfway to the first BZ edge. In this region, the anharmonic decay
presents a component of Umklapp processes, which is absent in the
vicinity of ${\bm \Gamma}$, where the scattering is almost entirely Normal.
On the other hand, the ZA broadening is relatively feature-less and
isotropic; it is quadratic in {\bf q} in the center of the first BZ then it
saturates and remains roughly constant.

\subsubsection{Phonon mean free path\label{sec_mfp}}
\begin{figure}
 {\includegraphics{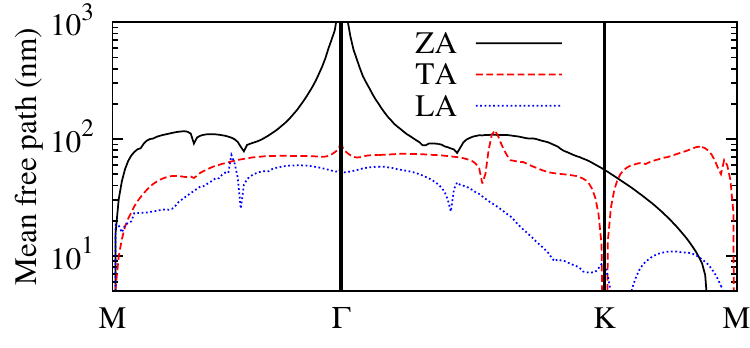}}%
\caption{\label{gene-mfp} (Color online) Phonon mean free path for the three acoustic branches in graphene at $300$~K.}
\end{figure}

\begin{figure*}[!tbp] 
\resizebox{1.80 \columnwidth}{!}
{\includegraphics{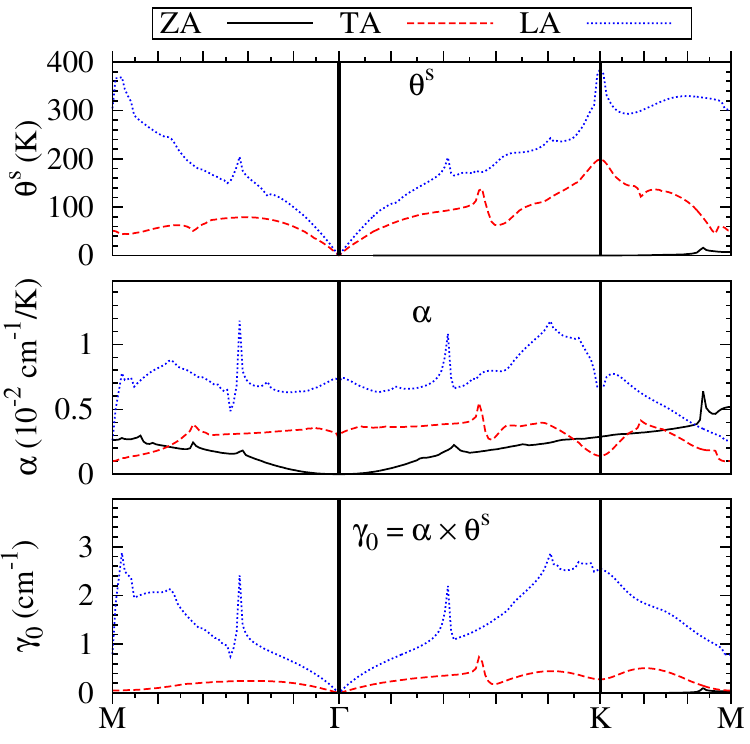}~\includegraphics{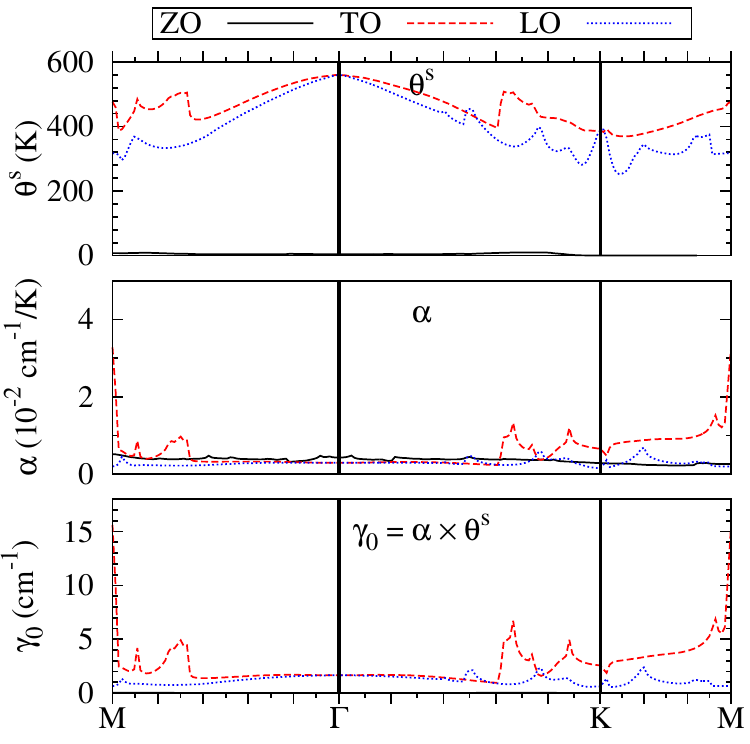}}
\caption{\label{gene-fit-A} 
(Color online)
$\theta^{\rm S}$ and $\alpha$ parameters (defined in the text) for the
graphene acoustic (left) and optical (right) branches.
For a given phonon mode, the temperature dependence of
the anharmonic broadening can be approximated by $\gamma(T)\simeq\alpha\theta^{\rm S}{\rm coth}(\theta^{\rm S}/T)$,
where $\theta^{\rm S}$ and $\alpha$ are the parameters corresponding
to that phonon.
$\theta^{\rm S}$ is a characteristic temperature and
$\alpha$ is the high temperature slope of $\gamma(T)$. $\gamma_0$ is
the $T=0$~K broadening.}
\end{figure*}

An alternative way to represent the effect of the phonon broadening is to plot the single-phonon mean free path (MFP):
\begin{align}
\lambda_{{\bf q}j}=\tau_{{\bf q}j}\left|{\bf c}_{{\bf q}j}\right|,
\end{align}
where ${\bf c}_{{\bf q}j}$ is the phonon group velocity and $\tau_{{\bf q}j}$ is the phonon lifetime, from equation \ref{eq_tauanh}.
In figure \ref{gene-mfp} we have plotted the MFP at $300$~K for the
three acoustic branches.
The MFP for the TO and LO bands is of order $100$~nm or smaller.
We have verified that it does not get substantially higher at lower temperatures, except
in the vicinity of ${\bm \Gamma}$ where it diverges at $0$~K.
At room temperature, the MFP of the ZA bands is one order of magnitude larger, i.e.
of order $1$~$\mu$m, in the center region of the Brillouin zone.
Furthermore, it increases as $1/T$ when temperature decreases.
Obviously, especially at small temperatures, when the intrinsic anharmonic MFP is too big,
other effects (typically the scattering on the borders of the sample) become important
and limit the value of the MFP.
We remark that the MFP of acoustical phonons is only one order of
magnitude smaller than the typical dimensions of high-quality graphene
samples.
It is, also, definitely larger than the transverse dimension of graphene nano-ribbons.
This result suggests that ballistic phonon-driven conductance could play a relevant role in this
kind of systems.

\subsubsection{Temperature dependence}

The intrinsic anharmonic
broadening of a specific phonon ({\bf q}j) has, in general, a typical dependence of the temperature $T$:
It is almost constant below a certain characteristic temperature $\theta^{\rm s}$,
then it rapidly becomes linear in $T$. Such a behavior is
reproduced by Eq.~\ref{eq_tauanh}.
A quadratic dependence on $T$ can be observed experimentally only at relatively high
$T$ and it is due to terms of order higher than those included in Eq.~\ref{eq_tauanh}.
~\cite{balkanski83}

From Eq.~\ref{eq_tauanh}, one can check that
\begin{equation}
\lim\limits_{T \to \infty} \gamma_{{\bf q}j}(T) = \alpha_{{\bf q}j}T + \mathcal{O}(1/T),
\end{equation}
where 
\begin{align}
\alpha_{{\bf q}j}={}&\frac{\pi K_{\rm B}}{\hbar^3 N_q}
\sum_{{\bf q'},j',j''}
\left|V^{(3)}_{{\bf q}j,{\bf q'}j',{\bf
      q''}j''}\right|^2\nonumber\\
&\times
\Bigg[
\left(\frac{1}{\omega_{{\bf q'}j'}}+\frac{1}{\omega_{{\bf q''}j''}}\right)
\delta(\omega_{{\bf q}j}-\omega_{{\bf q'}j'}-\omega_{{\bf
    q''}j''})\nonumber \\
&\phantom{\times}+2\left(\frac{1}{\omega_{{\bf q'}j'}}-\frac{1}{\omega_{{\bf q''}j''}}\right)
\delta(\omega_{{\bf q}j}+\omega_{{\bf q'}j'}-\omega_{{\bf q''}j''})
\Bigg]
\label{eq_alpha}
\end{align}
does not depend on $T$.
One can thus be tempted to approximate the overall dependence on
$T$ of the broadening $\gamma$ by
\begin{equation}
\gamma_{{\bf q}j}(T)\simeq
\tilde\gamma_{{\bf q}j}(T)=\alpha_{{\bf q}j}\theta^{\rm S}_{{\bf q}j}
~ {\rm coth} \left(\frac{\theta^{\rm S}_{{\bf q}j}}{T}\right),
\label{eq_tildegamm}
\end{equation}
where ${\rm coth}$ is the hyperbolic cotangent function, 
$\theta^{\rm S}_{{\bf q}j}=\gamma_{{\bf q}j}(0)/\alpha_{{\bf q}j}$, and
$\gamma_{{\bf q}j}(0)$ is the $T=0$ broadening from Eq.~\ref{eq_tauanh}.
Indeed, $\tilde\gamma$ from Eq.~\ref{eq_tildegamm}
is almost constant for $T\ll\theta^{\rm S}$ and tends to $\gamma(0)$
for $T\rightarrow0$.
Moreover, $\tilde\gamma(T) = \alpha T + \mathcal{O}(1/T)$
for $T\gg\theta^{\rm S}$.
To check the validity of this approximation (Eq.~\ref{eq_tildegamm}),
we systematically 
computed the graphene broadening for different phonons in the 
temperature range between 0 and 1500~K, using Eq.~\ref{eq_tauanh}.
These results are reasonably well reproduced by Eq.~\ref{eq_tildegamm}
with an error less than 5\%.

\begin{figure*}
 \resizebox{1.00 \textwidth}{!}{
 \includegraphics{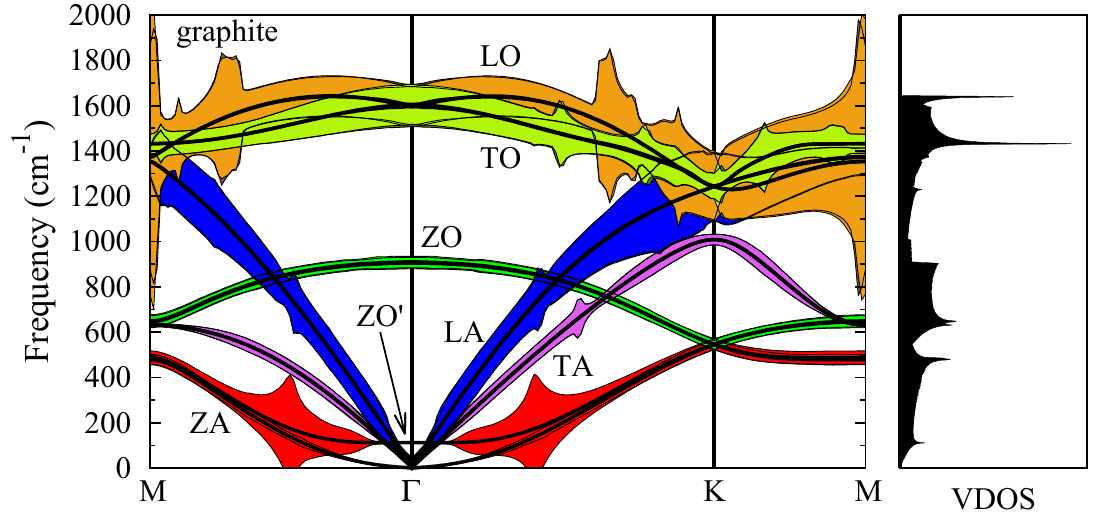}
 \includegraphics{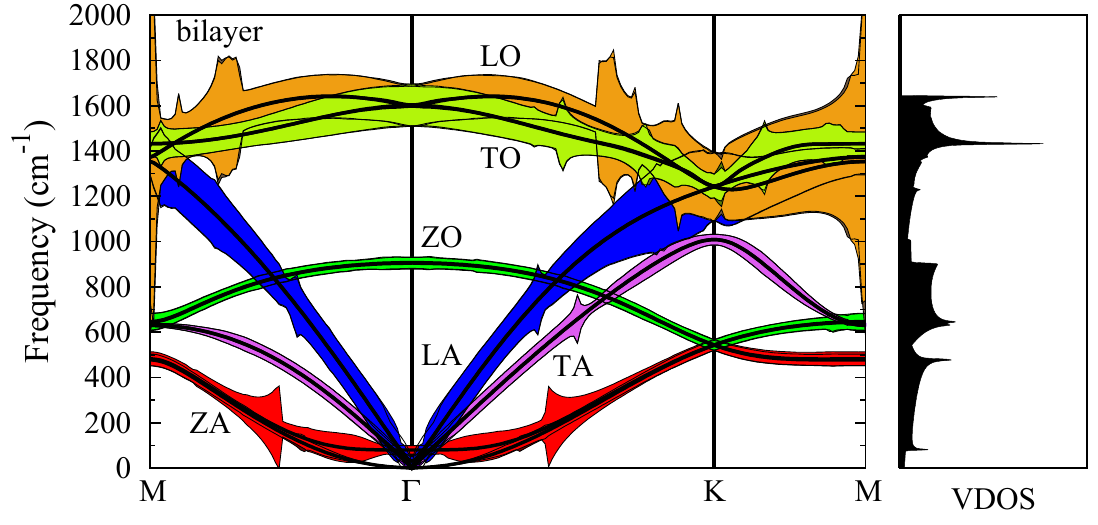}
  }
\caption{\label{gite-gn2l-disp} (Color online) Graphite and graphene
bilayer: calculated phonon dispersion widened by the anharmonic broadening (FWHM
$\times 100$) at 300 K.}
\end{figure*}

As a consequence, for a given phonon mode (${\bf q}j$), the knowledge of the two
corresponding parameters $\theta^{\rm S}$ and $\alpha$ is enough
to determine the overall temperature behavior of the broadening,
by using Eq.~\ref{eq_tildegamm}.
The two parameters can be extracted from Fig.~\ref{gene-fit-A},
for the graphene acoustic branches, along high symmetry lines.

Finally, the broadening of the LA and TA branches can be fitted with
an isotropic function of $q =|{\bf q}|$ and of $T$ of the form:
\begin{equation}
\gamma(q,T) = q B \coth\left(\frac{q A}{T}\right).
\end{equation}
By defining $b_0=\frac{2 \pi}{a_0}$, where $a_0$ is the cell parameter, we have:
for the LA band, $B_{\mathrm LA} = 4.58$~cm$^{-1}$$/b_0$ and $A_{\mathrm LA} = 694$~K$/b_0$;
for the TA band, $B_{\mathrm TA} = 0.805$~cm$^{-1}$$/b_0$ and $A_{\mathrm TA} = 241$~K$/b_0$.
These fitted parameters reproduce the computed linewidth with an error
of generally less than $10$\% for $q<0.40~b_0$ (smaller
for higher temperature and lower $|{\bf q}|$).
For the ZA band, $\theta^{\rm s}=0$ can be assumed, resulting in this simple fitting function:
\begin{equation}
\gamma(q,T) = B q^2 T\,.
\end{equation}
Taking $B_{\rm ZA} = 25.9\times10^{-3}$~cm$^{-1}$$/b_0^2$ we reproduce the value of linewidth to $10$\% accuracy in the range $q<0.25~b_0$, except for systematically underestimating it in the very small ${\bf q}$ region where it is negligible ($\gamma<10^{-5}$~cm$^{-1}$).

\subsection{Graphite and bilayer graphene}
\label{sec32}

We now discuss the anharmonic broadening in graphite and graphene bilayer.
Each of the six phonon branches of the graphene monolayer splits into
two branches for both graphite and graphene bilayer.
The three acoustic branches of graphene (ZA, TA, LA)
split into three acoustic (ZA, TA, LA) and three
quasi-acoustic branches (ZO$^\prime$, TO$^\prime$, LO$^\prime$).
The quasi-acoustic branches are almost degenerate with their respective
acoustic ones, except in the vicinity of the ${\bm \Gamma}$-{\bf A} line.
The remaining six optical branches are pair-wise quasi-degenerate in
the entire Brillouin zone, and they will be referred in pairs simply
as ZO, TO and LO or, in some cases as ZO$^1$, ZO$^2$, etc.
This notation does not hold along the ${\bm \Gamma}-{\bf A}$ line in graphite,
as degeneracy changes. It is however still possible to name
the branches by continuity below $600$~cm$^{-1}$.

\begin{figure}
 \resizebox{1.00 \columnwidth}{!}
{\includegraphics{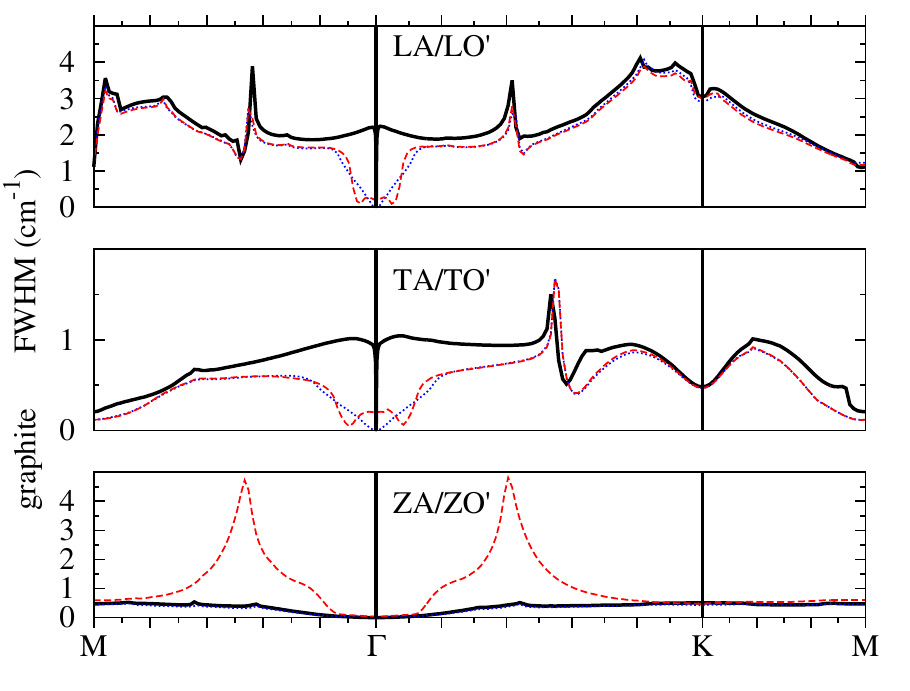}}
\caption{\label{gene-gite} (Color online) Anharmonic phonon broadening
(FWHM) at 300 K. Graphene acoustic branches (solid line) are compared
with the corresponding graphite acoustical (dotted) and quasi-acoustical
(dashed) ones.}
\end{figure}

\begin{figure}
 \resizebox{1.00 \columnwidth}{!}
{\includegraphics{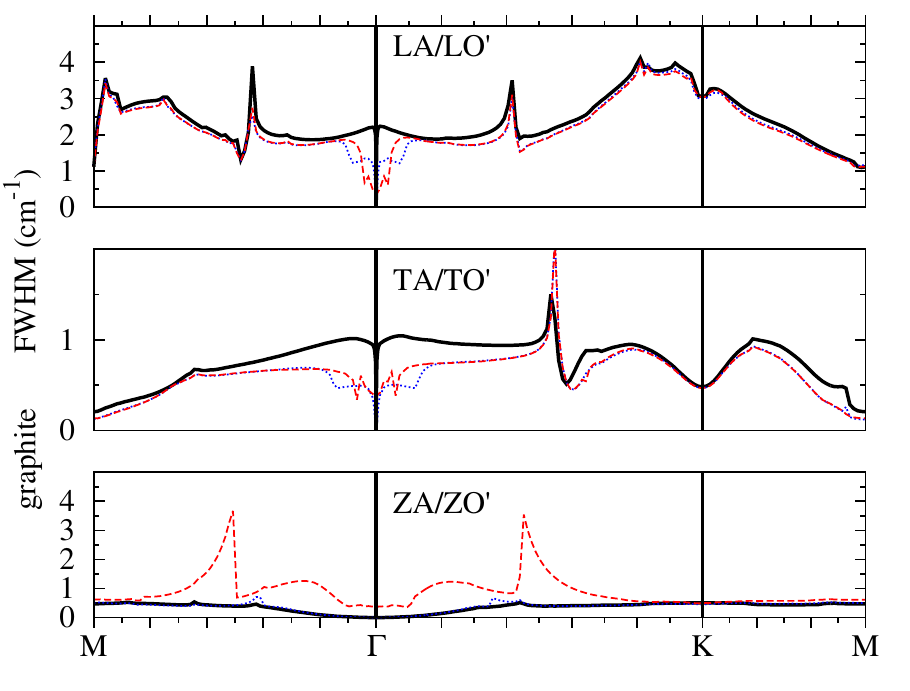}}%
\caption{\label{gene-gn2l} 
(Color online) Anharmonic phonon broadening
(FWHM) at 300 K. Graphene monolayer acoustic branches (solid line) are compared
with the corresponding bilayer acoustical (dotted) and quasi-acoustical
(dashed) ones.}
\end{figure}

\begin{figure*}
\resizebox{1.00 \textwidth}{!}{
 \includegraphics{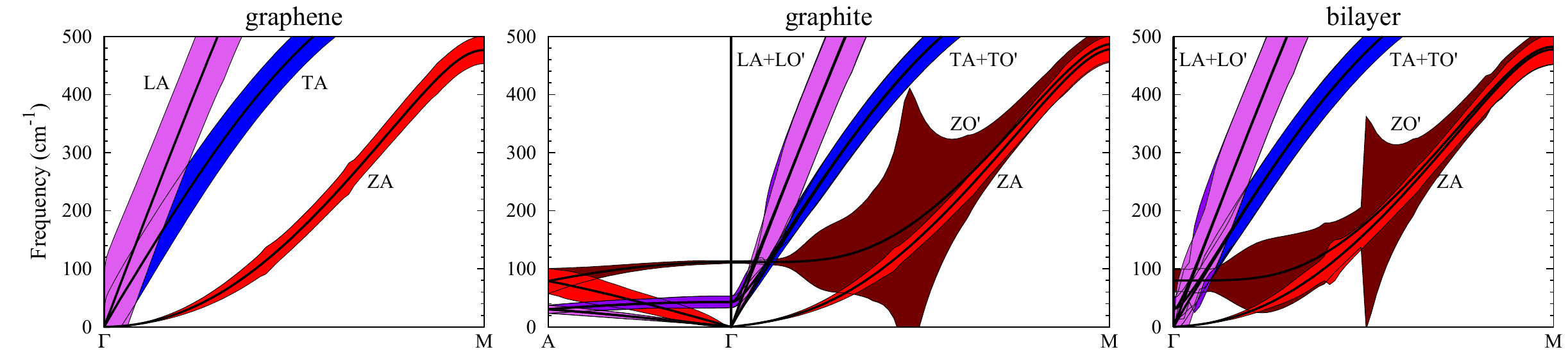}
 }%
 \\
%
\resizebox{1.00 \textwidth}{!}{
 \includegraphics{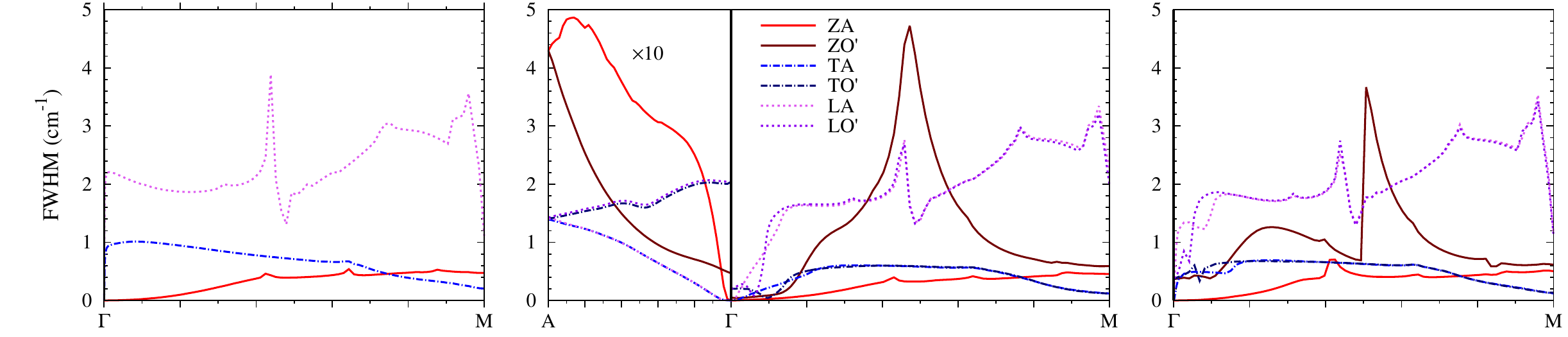}
 }%
\caption{\label{fig10} (Color online) 
Phonon dispersion and anharmonic broadening of the acoustic and
quasi-acoustic branches of graphene, graphite, and graphene bilayer.
The upper panels show the lower part of the phonon dispersion as in
Figs.~\ref{gene} and~\ref{gite-gn2l-disp}.
The lower panels show the FWHM at 300 K.
The values in the {\bf A}-${\bm \Gamma}$ section for graphite are magnified by a factor 10.}
\end{figure*}

Fig.~\ref{gite-gn2l-disp} shows a general view of the calculated phonon dispersion
and broadening in bulk graphite and graphene bilayer.
Fig.~\ref{gene-gite} and Fig.~\ref{gene-gn2l}, compare the broadening
of the acoustic and quasi-acoustic branches of, respectively, graphite
and bilayer graphene with those of the single layer graphene.
Fig.~\ref{fig10}, shows in more detail the low frequency region.
In the high energy part of the spectrum, graphite and graphene
monolayer and bilayer are almost indistinguishable, Fig.~\ref{gite-gn2l-disp},
meaning that, also in graphite, the physics is ruled by the two dimensional character of the phonon
dispersion.
Some of the graphene sharp features are a slightly broader in graphite due to the
out-of-plane phonon dispersion acting like an effective smearing.
The most striking differences between three-dimensional
bulk graphite and two-dimensional graphene monolayer and bilayer are associated with the
acoustic and quasi-acoustic branches.

First, we remind that, in graphene, the vibrational density of states (VDOS in Fig.~\ref{gene})
has a finite constant value for energies approaching zero.
This is because the ZA branch has a quadratic dispersion (not linear
as usual) and the VDOS is calculated in a two dimensional Brillouin zone.
On the contrary, in graphite, the VDOS goes to zero almost linearly
for energies going to zero, Fig.~\ref{gite-gn2l-disp}.
This happens in spite of the fact that in graphite, when the
out-of-plane component $q_{\rm z}=0$, the ZA branch is not very different
from the graphene one. In particular, on the scale of
Fig.~\ref{gite-gn2l-disp}, the graphite ZA branch (for $q_z=0$) appears quadratic as
the graphene one from Fig.~\ref{gene}.
However, for graphite, the VDOS is calculated in a three dimensional BZ
and the $q_z=0$ phonons have an infinitesimal weight.
Also, notice in the graphite VDOS, 
the presence of a peak at 132~cm$^{-1}$ corresponding to the
ZO$^\prime$ frequency at ${\bm \Gamma}$.
This peak and this phonon do not have a correspondence in graphene.
Finally, from Fig.~\ref{gite-gn2l-disp}, the bilayer VDOS has a
finite constant value for zero energy (as for the monolayer) and it
also shows the ZO$^\prime$ peak at 94~cm$^{-1}$ (as in graphite).

Let us consider the broadening of the LA and TA branches in Fig.~\ref{gene-gite}.
As already said, in graphene, these broadening do not vanish for $q\rightarrow 0$
and, for small $q$, they are relatively constant over a wide range
of $q$ values (\textit{e.g.} for the LA mode we are considering the
region with ${\bf q}<0.4{\bf M}$ and ${\bf q}<0.4{\bf K}$ in Fig.~\ref{gene-gite}).
This, ``plateau'' is due to Normal scattering towards 
the ZA phonons and its characteristics stem for the fact that the
ZA dispersions is quadratic and that the integration (the sum
in Eq.~\ref{eq_tauanh}) is done on a two dimensional Brillouin zone.
On the contrary, for three dimensional graphite, the broadening
of both LA and TA modes vanishes for $q\rightarrow 0$.
It is remarkable, however, that the graphite broadening still presents
a Normal scattering plateau, similar to the one of graphene, for
sufficiently large $q$. This is particularly evident for the LA mode
for ${\bf q}>0.2{\bf M}$ and ${\bf q}>0.2{\bf K}$ in Fig.~\ref{gene-gite}.
The LA and TA broadening in bilayer graphene, Fig.~\ref{gene-gn2l},
are rather more similar to the graphite one than to the graphene ones,
indicating that, for a higher number of layers, the broadening
should rapidly converge to the bulk graphite one.
Note that, graphene bilayer presents nonvanishing broadening of the TA
and LA bands at ${\bm \Gamma}$, its magnitude being about half for
bilayer than for graphene.

Finally, let us consider the $z$ polarized branches.
The ZA broadening in graphite is similar to the graphene one.
On the other hand, the ZO$^\prime$ broadening of graphite is much larger than the ZA one,
in spite of the fact that the ZO$^\prime$ and ZA branches are strictly related.
Particularly striking is the sudden increase in the ZO$^\prime$ broadening 
for certain values of {\bf q} in the vicinity of ${\bm \Gamma}$
(\textit{e.g} for ${\bf q}=0.47{\bf M}$ along the ${\bm \Gamma}{\bf M}$
direction). This peak of the broadening is due to the decay of a ZO$^\prime$ phonon, having
a finite wavevector {\bf q} into a ZA phonon near ${\bf q}$
and a LO$^\prime$ (or TO$^\prime$) phonon near ${\bm \Gamma}$.
This kind of decay is possible only when the energy difference between
the ZO$^\prime$ and the ZA is equal or smaller than the energy of the LO$^\prime$ and
TO$^\prime$ at ${\bm \Gamma}$ (the LO$^\prime$ and TO$^\prime$ are degenerate at
${\bm \Gamma}$), see Fig.~\ref{fig10}.
This condition is verified only for {\bf q} sufficiently far from
${\bm \Gamma}$. Thus, by increasing $q$, the sudden availability of this new decay channel
produces the peak in the broadening.
Finally, for the bilayer, the ZO$^\prime$ broadening presents a structure
which is not fundamentally different from the one in graphite.

\subsection{Thermal transport\label{sec_linew}}
\label{sec33}

\begin{figure*}
 \resizebox{1.00 \textwidth}{!}{
 \includegraphics{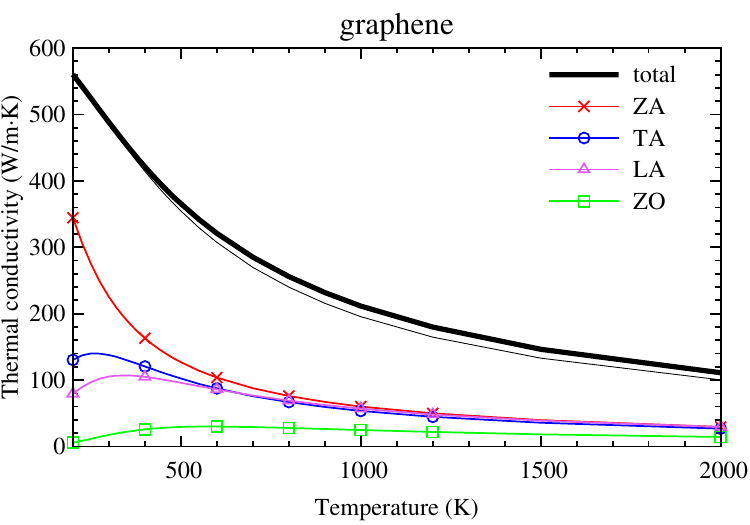}
 \includegraphics{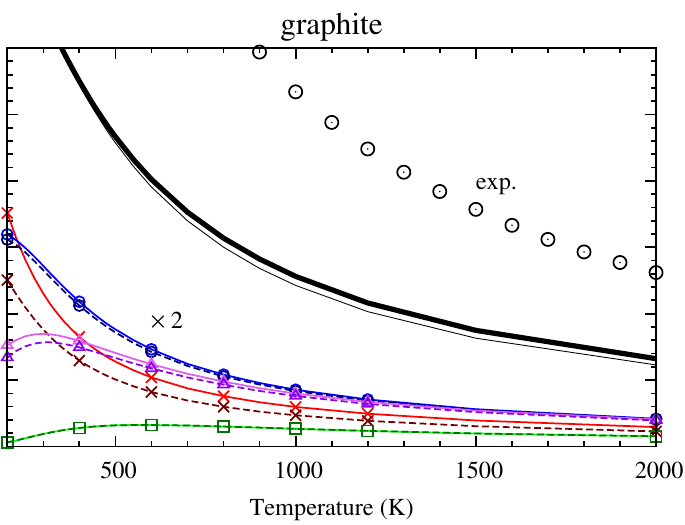}
 \includegraphics{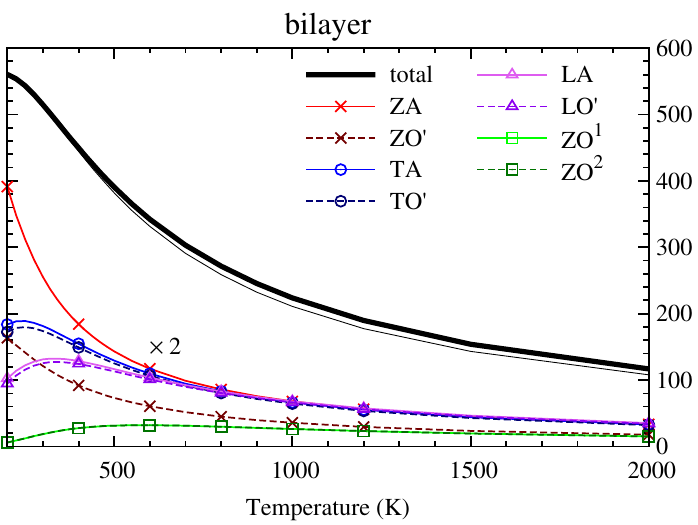}
}
\caption{\label{fig11} (Color online)
In-plane thermal conductivity obtained within SMA
(thick solid line) and its decomposition into the contributions
due to different phonon branches.
The figure shows only the contributions from a subset of the most relevant phonon branches.
The sum of these partial contributions is the thin solid line close to the total conductivity.
In order to compare more easily the curves, the single branch
contributions of graphite and bilayer are scaled by a factor 2.
Dots are experimental data from Ref.~\onlinecite{experiment}.}
\end{figure*}

The intrinsic anharmonic thermal conductivity $\kappa_{\rm L}$ has been computed
within the single mode time relaxation approximation
using Eq.~\ref{eq_k} .
For the two dimensional materials (graphene monolayer and bilayer) we
have used the convention that the volume $\Omega$ in Eq.~\ref{eq_k} is
the surface planar unit cell multiplied by the inter-layer distance of
graphite, $3.32$~\AA.
Fig.~\ref{fig11} reports the thermal conductivity and its
decomposition into different branch contributions.
In the temperature range considered, the conductivity is almost
entirely due to the acoustic and the quasi-acoustic branches.
Calculations of Fig.~\ref{fig11} are done for $T>200$~K.
Below that temperature, converged results can be obtained only
by using a much finer q-points grid than those presently used.

\begin{figure}
\resizebox{.4 \textwidth}{!}
{\includegraphics{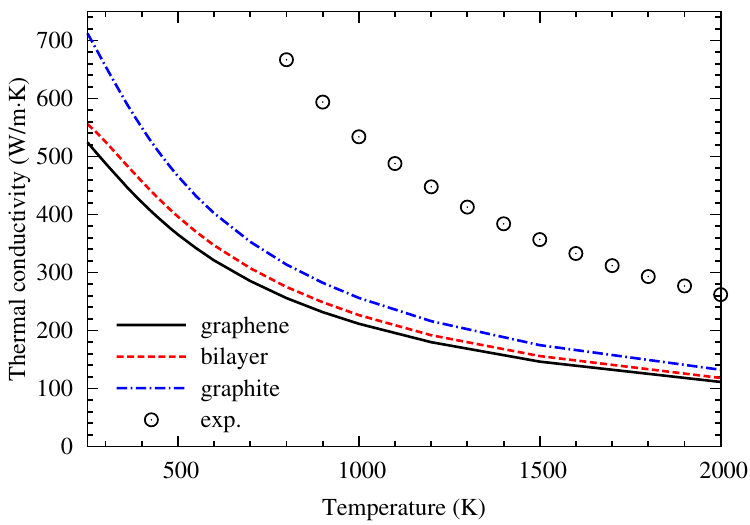}}
\caption{\label{fig12} (Color online) In-plane thermal conductivity calculated within
the SMA for for graphene single and bilayer and bulk graphite.
Measurements (exp.) are done on graphite and are from Ref.~\onlinecite{experiment}.}
\end{figure}

Let us, first, consider graphene. From Fig.~\ref{fig11}, the ZA contribution increases by
decreasing the temperature (actually, it
diverges for $T\rightarrow0$), while the LA and TA contributions
are non monotonic and reach a maximum near 300 K.
The difference in the two behaviors can be understood by considering that, for
small $T$, the phonons mostly occupied have small $q$, and that,
for $q\rightarrow 0$, the anharmonic broadening 
(the inverse of the $\tau$ appearing on the r.h.s of Eq.~\ref{eq_k})
of the ZA mode goes to zero at any $T$.
This is not the case for
the broadening of the TA and LA branches, Fig.~\ref{gene2} .
Now, let us compare in Fig.~\ref{fig11} graphene with graphite.
The ZA contribution in graphene corresponds, in graphite, to the two separate
contributions ZA and ZO$^\prime$. These two are quite different already
at room temperature. Below room temperature, the ZA increases and diverges for
$T\rightarrow0$ (as for the ZA in graphene), while the ZO$^\prime$ does not.
The TA contribution in graphene corresponds, in graphite, to the
TA and TO$^\prime$ ones. Above 200~K, the TA and TO$^\prime$ contributions
are almost indistinguishable, in spite of the fact that only the TA
branch is actually acoustical.
Important differences between TA and TO$^\prime$ contributions appear only
below 50~K (not shown). The same considerations hold for the LA LO$^\prime$ couple.
By comparing in Fig.~\ref{fig11} graphite with the bilayer, above
200~K, the overall behavior of the two systems is relatively
similar, in spite of the different dimensionality.
Indeed, in the same temperature range, the total conductivity of 
two dimensional graphene mono- and bi-layer and that of three
dimensional graphite are relatively very similar, Fig.~\ref{fig12},
with a difference between graphene and graphite of less than 10\%.

Fig.~\ref{fig11} also reports the measured in-plane thermal conductivity.
This is done only for bulk graphite, because of the abundance of experimental data.
At present, experimental measurements on graphene exist only for small samples, 
where border effects are important, consequently the range of measured values is
large and the number of samples in temperature limited\cite{balandin}.
From Fig.~\ref{fig11}, the calculated graphite conductivity (which is
obtained  within the SMA) grossly underestimates the measured one, by about
a factor two.
It is unlikely that this disagreement is due to density functional
theory. Indeed, DFT reproduces accurately the measured graphite phonon
dispersion, Fig.~\ref{fig0}, suggesting that the most
important quantities used in Eq.~\ref{eq_k} are correct.
On the other hand, as already explained at the end of
Sec.~\ref{sec_anh}, the thermal conductivity calculated according
to Eq.~\ref{eq_k} derives from the single mode relaxation
time approximation and not from an exact solution of the transport equation.
Indeed, according to Refs.~\onlinecite{mingo-gite,mingo10}, the
SMA cannot be used to properly describe the in-plane thermal conductivity in graphitic
materials and the exact solution of the Boltzmann transport equation
is required. However, the results of Refs.~\onlinecite{mingo-gite,mingo10} 
are obtained by using a semi-empirical interatomic potential and
a direct comparison with the present results is not meaningful.
Further investigation on this point is required.

\begin{figure}
\resizebox{.4 \textwidth}{!}
 {\includegraphics{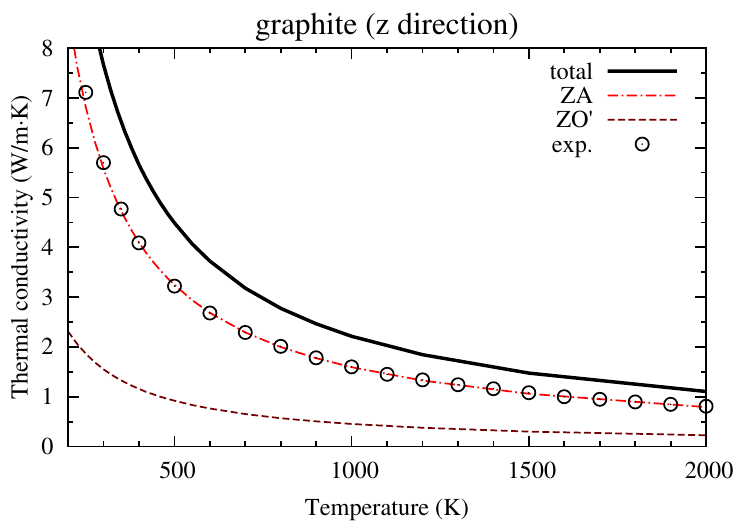}}%
\caption{\label{fig13} (Color online) Thermal conductivity of graphite along the direction orthogonal to the planes
and its decomposition into different phonon contributions.
Measurements (exp.) are from Ref.~\onlinecite{experiment} and should be compared with the line labeled as ``total''.
}
\end{figure}

We now consider the thermal conductivity
along the $z$ axis, perpendicular to the graphene planes.
Fig.~\ref{fig13} shows calculations and compares them with measurements.
The quasi totality of the conduction is due to the acoustic and quasi-acoustic
phonons polarized along $z$ and, as expected, the conduction is much smaller than the in-plane one.
The transport in-plane and the one along $z$ are quite different.
The phonons relevant for the $z$ conduction
have much smaller velocity and are localized around $\gamma$ in reciprocal space. Indeed, 
only phonons with a nonzero dispersion along the $z$-axis can conduct
since they have a nonzero velocity and, thus, can give a contribution to the
sum in Eq.~\ref{eq_k}. These phonons belong a small region surrounding the ${\bm \Gamma}$-{\bf A} line. If we only integrate the contribution from a z$z$-oriented {\bf q}-space cilinder centered around $\Gamma$, we estimate that, at 300~,K the central 20\% of the WS cell contributes about 85\% of the transverse transport, but only aout 40\% of the in-plane transport.

We also remark that the the $z$ conduction is extremely sensitive to the value of the $c/a$
lattice parameter.
Indeed, a small change in $c/a$ results in a systematic increase or decrease
of the frequencies of all the phonons relevant for the transport along $z$, see Fig.~\ref{fig0}.
Moreover, a systematic rescale of phonon frequencies by a certain factor $\lambda$
results in a rescale of the conduction by a factor which can be much bigger
than the initial $\lambda$ (see Eq.~\ref{eq_k}).
The agreement with measurements from Fig.~\ref{fig13} is satisfactory; the theoretical calculations slightly overrestimates the experimental value, as we expect since we are omitting isotopic effects.
We judge it compatible with the assumption that the SMA correctly describes the
thermal transport along the $z$ axis.

\section{Conclusions}
\label{sec4}

We have developed and implemented a generic method for the calculation of
anharmonic three-phonon scattering coefficients within
density functional theory.
The three phonons can have three arbitrary wavevectors $({\bf q}, {\bf q'}, {\bf q''})$.
The approach works for materials with an electronic gap and also for
metallic or zero gap materials.
The method has been implemented in the {\sc Quantum ESPRESSO} package~\cite{wwwqe,*qe}
and generalizes a previously existing code developed in Ref.~\onlinecite{lazzeri02}.
The anharmonic coefficients which can be obtained can be used in a straightforward way to
compute the anharmonic broadening of a phonon with an arbitrary
wavevector {\bf q} and the intrinsic thermal conductivity within
the single-mode relaxation time approximation (SMA).
The first applications have been devoted to study graphite, graphene mono- and bi-layer.

We have reported a detailed analysis of the anharmonic phonon broadening in graphene.
Interesting, the broadening of the high-energy optical branches is highly nonuniform
and presents a series of sudden steps and spikes of various origin.
At finite temperature, the two linearly dispersive acoustic branches TA and LA 
have nonzero broadening for $q\rightarrow 0$, as noticed in Ref.~\onlinecite{bonini12}.
This anomalous behavior is due to Normal scattering towards two ZA phonons
(which are quadratically dispersive), for small $q$.
The activation of the Umklapp scattering for sufficiently large $q$
is associated with a sudden increase of the broadening at nearly
half the Wigner Seitz cell.
We provide a set of expressions which ca be used to fit the anharmonic
scattering time and broadening for the acoustic phonon branches,
which are the most relevant in thermal transport.

The broadening of graphite and bi-layer is, overall, very similar to the graphene one.
The most remarkable feature is the broadening of the quasi acoustical ZO$'$ branch,
which is much larger and very different 
from the one of the strictly related ZA acoustic branch.
On the other hand, the broadening of the TA and LA branches
of graphite, displays a certain number of similarities
with that of graphene mono and bi-layer,
in spite of the different dimensionality of the systems.

Finally, we have calculated the intrinsic anharmonic thermal conductivity
within the SMA.
The in-plane conductivity in graphite, graphene mono- and bi-layer 
are very similar in spite of the differences among these systems.
The calculated SMA conductivity heavily underestimates the measured
one  (for graphite) by almost a factor two, in the temperature range from
200 to 2000 K. This finding is compatible with the conclusions of
Ref.~\onlinecite{mingo-gite,mingo10}, which state that the SMA cannot be
used to properly describe the in-plane thermal conductivity in graphitic
materials.
On the other hand, the calculated SMA conductivity for graphite along
the direction perpendicular to the plane is in good agreement
with measurements.

\section* {Acknowledgments}
This work was financed by ANR project \textsc{accattone}.
Calculations were done at IDRIS (Orsay, France), 
Project No.~096128 and CINES Project \emph{imp6128}.
We thank G. Fugallo for discussions.

\appendix

\section{Third order calculation}
\label{app1}

This section describes the method used to calculate the third order anharmonic
scattering coefficients. 
In order to fix the notation, Sec.~\ref{seca1}, and ~\ref{seca2}
resume DFT and linear response to DFT (alias DFPT).
Third order calculations are described in Sec.~\ref{seca3}.
Sec.~\ref{seca4}, and ~\ref{seca5} gives the explicit expressions for certain terms.
Sec.~\ref{seca6} describes the implementation of the nonlinear core corrections.

\subsection{Kohn-Sham equations}
\label{seca1}
Within DFT the total energy a system can be determined from the ground
state electronic charge density $n$.
In turn, $n$ can be obtained by solving self-consistently
the Kohn Sham (KS) equations\cite{martin-book}, which, in a periodic crystal are:
\begin{align}
&[T^{\rm kin}+V^{\rm KS}]|\psi_{{\bf k},i}\rangle=\epsilon_{{\bf k},i}~|\psi_{{\bf k},i} \rangle
\label{eqks1}\\
&V^{\rm KS}({\bf r})=v^{\rm ion}({\bf r})+\frac{\delta E_{\rm
    I}[{n}]}{\delta n({\bf r})}
\label{eqks2}\\
&n({\bf r})=\sum_{{\bf k},i} 
\tilde\theta_{{\bf k},i}
|\langle\psi_{{\bf k},i} |{\bf r}\rangle|^2
~;~\int n({\bf r}) d{\bf r} = N^{\bf el}
\label{eqks3}
\end{align}
In Eq.~\ref{eqks1},
$T^{\rm kin}$ is the single-particle kinetic energy operator,
$V^{\rm KS}$ is the KS potential,
$|\psi_{{\bf k},i}\rangle$ 
are the Bloch eigenstates with wavevector {\bf k},
band index $i$, and energy $\epsilon_{{\bf k},i}$.
$\langle {\bf r}+{\bf R}|\psi_{{\bf k},i}\rangle=e^{i{\bf k}\cdot{\bf R}}
\langle {\bf r}|\psi_{{\bf k},i}\rangle$, being
{\bf r} the position and {\bf R} a lattice vector.
$v^{\rm ion}$ is the external potential
due to the ions, $E_{\rm I}[n]$ is the interaction functional
(Hartree energy plus exchange-correlation contribution).
The sum in Eq.~\ref{eqks3} is done on a sufficiently fine grid of
{\bf k}-points.
$\tilde\theta_{{\bf k},i}$ is the occupation of an electronic state:
$\tilde\theta_{{\bf k},i}=1$ for valence band electrons and
$\tilde\theta_{{\bf k},i}=0$ for conduction ones.
Here and in the following $\int d{\bf r}$ is the integral
over all the space. $N^{\rm el}$ is the total number of valence electrons
(we consider $e^2=1$).
The total energy of the system is:
\begin{equation}
{\cal E}^{\rm tot}=E^{\rm ion}+\sum_{{\bf k},i}\epsilon_{{\bf
    k},i}\tilde\theta_{{\bf k},i}+E_{\rm I}[n]-\int \frac{\delta
  E_{\rm I}[n]}{\delta n({\bf r})} n({\bf r})d{\bf r},
\label{eqen}
\end{equation}
where $E^{\rm ion}$ is the ionic contribution.

In this form, the KS equations are suitable 
for semiconductor or insulators, where the electronic gap is different from zero.
if the electronic gap vanishes (metal and semi-metal case) it is customary~\cite{methfessel89}
to introduce a
smearing function $\theta_\sigma(x)$, which is characterized by a
smearing width $\sigma$, and which becomes a step function in
the limit $\sigma\rightarrow0$.
The KS equation are still written as in Eqs.~\ref{eqks1}-\ref{eqks3},
but now $\tilde\theta_{{\bf
    k},i}=\theta_\sigma(\epsilon_{\rm F}-\epsilon_{{\bf k},i})$, where the
Fermi energy $\epsilon_{\rm F}$ has to be determined self consistently from
\begin{equation}
\sum_{{\bf k},i}\tilde\theta_{{\bf k},i} = N^{\rm el}.
\end{equation}
Furthermore, in the metallic case, a proper definition of the energy
${\cal E}^{\rm tot}$ requires~\cite{degironcoli95}
Eq~\ref{eqen} to include the term 
\begin{equation}
\sum_{{\bf k},i}\int_{-\infty}^
{\epsilon_{\rm F}-\epsilon_{{\bf k},i}}
x\delta_\sigma(x)~dx,
\end{equation}
where $\delta_\sigma(x)=\partial\theta_\sigma(x)/(\partial x)$.

\subsection{Linear response}
\label{seca2}

The derivative of the electronic charge distribution with respect
to the {\bf q} periodic displacement, $u_{\bf q}$ as defined in Eq.~\ref{equq}
(for simplicity from now on we will drop the indexes $\alpha$ and $s$),
can be obtained from first order perturbation theory~\cite{dfpt}.
For the metallic case, linear response can be implemented following
Ref.~\onlinecite{degironcoli95}; we will follow the equivalent, but slightly different
approach of Ref.~\onlinecite{lazzeri02}.

Let us consider a uniform grid of electronic {\bf k} points and a phonon 
wavevector {\bf q} which belongs to this grid.
First, one has to solve the KS equations and obtain the ground state
change density $n$ and the corresponding KS wavefunctions $|\psi_{{\bf
  k},i}\rangle$.
Then, one has to define a ``cutoff'' energy $\overline{E}$
which separates the electronic states which are completely empty
from those which are occupied or partially occupied.
In the semiconductor/insulator case, $\overline{E}$ can be any energy within the gap;
in the metallic case any $\overline{E}\ge
\epsilon_{\rm F}+3\sigma$ is a reasonable choice.
We define $P_c$ as the projector on the manifold spanned by the 
empty states and $P_v=1-P_c$ as the projector onto the occupied and
partially occupied states.

The derivative of the charge $\partial n/\partial u_{\bf q}$ and the
derivative of the KS wavefunctions projected onto the conduction bands
$|\phi^{{\bf q}}_{{\bf k},i}\rangle=P_c|\partial\psi_{{\bf
    k},i}/\partial u_{\bf q}\rangle$
can be obtained by solving self-consistently the equations:
\begin{equation}
[T^{\rm kin}+V^{\rm KS}+\alpha P_v -\epsilon_{{\bf k},i}]
|\phi^{\bf q}_{{\bf k}i}\rangle=
-P_c\frac{\partial V_{\rm KS}}{\partial u_{\bf q}}
|\psi_{{\bf k}i}\rangle
\label{eqlr1}
\end{equation}
\begin{equation}
\frac{\partial V^{\rm KS}({\bf r})}{\partial u_{\bf q}}=
\frac{\partial v^{\rm ion}({\bf r})}{\partial u_{\bf q}}+
\int
\frac{\delta^2E_{\rm I}[n]}{\delta n({\bf r})\delta n({\bf r'})}
\frac{\partial n({\bf r'})}{\partial u_{\bf q}} d{\bf r'}
\label{eqlr2}
\end{equation}
{\belowdisplayskip=0.0pt 
\begin{align}
\frac{\partial n({\bf r})}{\partial u_{\bf q}}&=
\sum_{{\bf  k}}
\langle{\bf r}|
\bigg\{
\sum_{i}
\tilde\delta_{{\bf k}i}
\epsilon^{\bf q}_{\rm F}
|\psi_{{\bf k}i}\rangle
\langle\psi_{{\bf k}i}| 
\nonumber \\
&+\sum_{i,j}^v 
\frac{ \tilde\theta_{{\bf k}i} - \tilde\theta_{{\bf k+q},j}}
{\epsilon_{{\bf k}i} - \epsilon_{{\bf k+q},j}}
|\psi_{{\bf k+q},j}\rangle
\langle \psi_{{\bf k+q},j}
|V^{\bf q}|
\psi_{{\bf k}i} \rangle
\langle\psi_{{\bf k}i}|
\nonumber \\
&+
\sum_i \tilde\theta_{{\bf k}i} \Big[
|\phi^{{\bf q}}_{{\bf k}i}\rangle\langle
\psi_{{\bf k}i}|+
|\psi_{{\bf k}i}\rangle\langle
\phi^{{\bf -q}}_{{\bf k}i}|
 \Big]
\bigg\}|{\bf r}\rangle
\label{eqlr3}
\end{align}
}
$\alpha$ is a constant chosen in such a way that the linear
system of Eq.~\ref{eqlr1} is not singular.
$\sum^v$ indicates that the sum is to be performed only on the
partially occupied states.
$\tilde\delta_{{\bf k},i}=\delta_\sigma(
\epsilon_{\rm F}-\epsilon_{{\bf k},i})$

The first two terms in the right hand side of Eq.~\ref{eqlr3}
are different from zero only in the metallic case
and are written using the notation
$\epsilon_{\rm F}^{\bf q}=\partial\epsilon_{\rm F}/\partial u_{\bf q}$ and
$V^{\bf q}=\partial V_{\rm KS}/\partial u_{\bf q}$.
$\epsilon_{\rm F}^{\bf q}=0$ for {\bf q}$\ne${\bf 0};
it has to be determined self-consistently from 
\begin{equation}
\epsilon^{\bf q}_{\rm F}=\frac{
\sum_{{\bf k},i} \langle\psi_{{\bf k},i}|V^{\bf q}|\psi_{{\bf
    k},i}\rangle}
{\sum_{{\bf k},i}\tilde\delta_{{\bf k},i}}.
\end{equation}

In the second line of Eq.~\ref{eqlr3} we have used a compact notation which reads:
when the denominator is equal to zero one has to substitute
the ratio with its limit as the denominator approaches zero.
The same substitution can be used when the denominator is
very small in order to gain numerical stability.
Thus, when 
$\epsilon_{{\bf k}i} \sim \epsilon_{{\bf k+q},j}$ one can substitute
$(\tilde\theta_{{\bf k}i} - \tilde\theta_{{\bf k+q},j})
/(\epsilon_{{\bf k}i} - \epsilon_{{\bf k+q},j})$
with $-\tilde\delta_{{\bf k},i}$.
We, finally, note that the present approach is different from the the one
described in Ref.~\onlinecite{degironcoli95} and that the $|\phi\rangle$
wavefunctions presently defined are different from the $|\phi\rangle$
of Ref.~\onlinecite{degironcoli95}.


\subsection{Third order}
\label{seca3}

Let us consider three phonon displacements 
$u_{{\bf q}}$, $u_{{\bf q}'}$, $u_{{\bf q}''}$ such that the sum of their wavevectors is
${\bf q}+{\bf q}'+{\bf q}''={\bf 0}$.
By solving the linear response equations one can obtain
$\partial n/\partial u_{{\bf q}}$ and the $\{|\phi^{\bf q}_{{\bf k},i}\rangle\}$
corresponding to the three phonons.
The third derivative of the energy can then be obtained from
\begin{widetext}
\begin{equation}
\frac{\partial^3 {\cal E}^{\rm tot}}{\partial u_{{\bf q}}\partial u_{{\bf q}'}\partial u_{{\bf q}''}}=
\frac{1}{6} \left[
\tilde E^{{\bf q}, {\bf q}', {\bf q}''} +
\tilde E^{{\bf q}'', {\bf q}, {\bf q}'} +
\tilde E^{{\bf q}', {\bf q}'', {\bf q}} +
\tilde E^{{\bf q}, {\bf q}'', {\bf q}'} +
\tilde E^{{\bf q}'', {\bf q}', {\bf q}} +
\tilde E^{{\bf q}', {\bf q}, {\bf q}''}
\right] \label{eqthird}
\end{equation}
\begin{equation}
\begin{split}
\tilde E^{{\bf q}, {\bf q}', {\bf q}''} = {} &
Z^{{\bf q}, {\bf q}', {\bf q}''}+
\frac{\partial^3 E^{\rm ion}}{\partial u_{{\bf q}}\partial u_{{\bf q}'}\partial u_{{\bf q}''}}+
\int 
n({\bf r})
\frac{\partial^3 v^{\rm ion}({\bf r})}{\partial u_{{\bf q}}\partial u_{{\bf q}'}\partial u_{{\bf q}''}}
d{\bf r}
\\
&+3\int
\frac{\partial n({\bf r})}{\partial u_{{\bf q}}}
\frac{\partial^2 v^{\rm ion}({\bf r})}{\partial u_{{\bf q}'}\partial u_{{\bf q}''}}
d{\bf r}
+\iiint
\frac{\delta^3 E_{\rm I}[n]}{\delta n({\bf r})\delta n({\bf r'})\delta n({\bf r''})}
\frac{\partial n({\bf r})}{\partial u_{{\bf q}}}
\frac{\partial n({\bf r'})}{\partial u_{{\bf q}'}}
\frac{\partial n({\bf r''})}{\partial u_{{\bf q}''}}
d{\bf r}d{\bf r'}d{\bf r''}
\label{eqa12}
\end{split}
\end{equation}

For the semiconductor/insulator case, we can follow Ref.~\onlinecite{debernardi94} and write:
\begin{equation}
Z^{{\bf q}, {\bf q}', {\bf q}''}=
6\sum_{{\bf k}}
\left[
\sum_i
\tilde\theta_{{\bf k}i}
\langle\phi^{-{\bf q}}_{{\bf k}i}|V^{{\bf q}'}|\phi^{{\bf q}''}_{{\bf k}i}\rangle
-\sum_{i,j}^v
\tilde\theta_{{\bf k}i}
\langle\phi^{-{\bf q}}_{{\bf k+q},i}|\phi^{{\bf q}'}_{{\bf k-q}',j}\rangle
\langle\psi_{{\bf k-q}',j}|V^{{\bf q}''}|\psi_{{\bf k+q},i}\rangle
\right] \label{eq_ins}
\end{equation}

\belowdisplayskip=15pt
For the metallic case, we follow Ref.~\onlinecite{lazzeri02}:
\begin{equation}
\begin{split}
Z^{{\bf q}, {\bf q}', {\bf q}''}={}&
\sum_{{\bf k}}\Bigg\{
6\sum_i
\tilde\theta_{{\bf k}i}
\langle\phi^{-{\bf q}}_{{\bf k}i}|V^{{\bf q}'}|\phi^{{\bf q}''}_{{\bf k}i}\rangle
 \\
&+
6\sum_{i,j}^v
\frac{
\left[
\tilde\theta_{{\bf k+q},i}
\langle\phi^{-{\bf q}}_{{\bf k+q},i}|V^{{\bf q}'}|\psi_{{\bf k-q}',j}\rangle
-
\tilde\theta_{{\bf k-q}',j}
\langle\psi_{{\bf k+q},i}|V^{{\bf q}}|\phi^{{\bf q}'}_{{\bf
    k-q}',j}\rangle
\right]
\langle\psi_{{\bf k-q}',j}|V^{{\bf q}''}|\psi_{{\bf k+q},i}\rangle
}{
\epsilon_{{\bf k+q},i}-\epsilon_{{\bf k-q}',j}
}  \\
&+
2\sum_{i,j,l}^v\Bigg[
\langle\psi_{{\bf k},i}|V^{{\bf q}}|\psi_{{\bf k-q},j}\rangle
\langle\psi_{{\bf k-q},j}|V^{{\bf q}'}|\psi_{{\bf k+q}'',l}\rangle
\langle\psi_{{\bf k+q}'',l}|V^{{\bf q}''}|\psi_{{\bf k},i}\rangle
 \\
&
\phantom{+2\sum_{i,j,l}^v\Bigg[}
\times\frac{
\tilde\theta_{{\bf k},i}
(\epsilon_{{\bf k-q},j}-\epsilon_{{\bf k+q}'',l})
+\tilde\theta_{{\bf k-q},j}
(\epsilon_{{\bf k+q}'',l}-\epsilon_{{\bf k},i})
+\tilde\theta_{{\bf k+q}'',l}
(\epsilon_{{\bf k},i}-\epsilon_{{\bf k-q},j})
}{
(\epsilon_{{\bf k},i}-\epsilon_{{\bf k-q},j})
(\epsilon_{{\bf k-q},j}-\epsilon_{{\bf k+q}'',l})
(\epsilon_{{\bf k+q}'',l}-\epsilon_{{\bf k},i})
}\Bigg]
\\
&+
3\epsilon^{{\bf q}}_{\rm F}\left[
\sum_{i,j}^v
\frac{\tilde\delta_{{\bf k},i}-\tilde\delta_{{\bf k+q}'',j}} {\epsilon_{{\bf k},i}-\epsilon_{{\bf k+q}'',j}}
\langle\psi_{{\bf k},i}|V^{{\bf q}'}|\psi_{{\bf k+q}'',j}\rangle
\langle\psi_{{\bf k+q}'',j}|V^{{\bf q}''}|\psi_{{\bf k},i}\rangle+
2\sum_i \tilde\delta_{{\bf k},i}
\langle\psi_{{\bf k},i}|V^{{\bf q}'}|\phi^{{\bf q}''}_{{\bf k},j}\rangle
\right]
\\
&+
3\epsilon^{{\bf q}}_{\rm F}\epsilon^{{\bf q}'}_{\rm F}
\left(\sum_i \tilde\delta_{{\bf k},i}^{(1)} 
\langle\psi_{{\bf k},i}|V^{{\bf q}''}|\psi_{{\bf k},i}\rangle
\right)
-\epsilon^{{\bf q}}_{\rm F}\epsilon^{{\bf q}'}_{\rm F}\epsilon^{{\bf q}''}_{\rm F}
\left(\sum_i \tilde\delta_{{\bf k},i}^{(1)} \right) 
\Bigg\},
\label{eq_zmet}
\end{split}
\end{equation}
where 
$\tilde\delta^{(1)}_{{\bf k},i}=\left.\partial
  \delta_\sigma(x)/(\partial x)\right|_
{x=\epsilon_{\rm F}-\epsilon_{{\bf k},i}}$.
Eq.~\ref{eq_zmet} is written with the same compact notation of \ref{eqlr3}:
when one of the denominators in Eq.~\ref{eq_zmet} vanishes the corresponding
term is replaced with its limit.
In particular, when $\epsilon_{{\bf k+q},i}\sim\epsilon_{{\bf k-q}',j}$
in the second line of Eq.~\ref{eq_zmet}, the argument of the sum 
can be written as
\[
-\left[
\tilde\theta_{{\bf k+q},i}
\langle\phi^{-{\bf q}}_{{\bf k+q},i}|\phi^{{\bf q}'}_{{\bf k-q}',j}\rangle
+\tilde\delta_{{\bf k+q},i}
\langle\psi_{{\bf k+q},i}|V^{{\bf q}}|\phi^{{\bf q}'}_{{\bf  k-q}',j}\rangle
\right]
\langle\psi_{{\bf k-q}',j}|V^{{\bf q}''}|\psi_{{\bf k+q},i}\rangle.
\]
\end{widetext}
The limits of the factor in the fourth line of Eq.~\ref{eq_zmet} are
\begin{align}
{\rm for}&~
\epsilon_{{\bf k},i}\sim\epsilon_{{\bf k-q},j}\ne\epsilon_{{\bf
    k+q}'',l}: \nonumber \\
&
\left[
\frac{\tilde\theta_{{\bf k},i}-\tilde\theta_{{\bf k+q}'',l}}
{\epsilon_{{\bf k},i}-\epsilon_{{\bf k+q}'',l}}
+\tilde\delta_{{\bf k},i}
\right]
\frac{1}{\epsilon_{{\bf k+q}'',l}-\epsilon_{{\bf k},i}}
\nonumber\\
{\rm for}&~
\epsilon_{{\bf k-q},j}\sim\epsilon_{{\bf k+q}'',l}\ne\epsilon_{{\bf k},i}:
\nonumber\\
&
\left[
\frac{\tilde\theta_{{\bf k-q},j}-\tilde\theta_{{\bf k},i}}
{\epsilon_{{\bf k-q},j}-\epsilon_{{\bf k},i}}
+\tilde\delta_{{\bf k-q},j}
\right]
\frac{1}{\epsilon_{{\bf k},i}-\epsilon_{{\bf k-q},j}
}
\nonumber\\
{\rm for}&~
\epsilon_{{\bf k+q}'',l}\sim\epsilon_{{\bf k},i}\ne\epsilon_{{\bf k-q},j}:
\nonumber\\
&
\left[
\frac{\tilde\theta_{{\bf k+q}'',l}-\tilde\theta_{{\bf k-q},j}}
{\epsilon_{{\bf k+q}'',l}-\epsilon_{{\bf k-q},j}}
+\tilde\delta_{{\bf k+q}'',l}
\right]
\frac{1}{\epsilon_{{\bf k-q},j}-\epsilon_{{\bf k+q}'',l}
}
\nonumber\\
{\rm for}&~
\epsilon_{{\bf k},i}\sim\epsilon_{{\bf k-q},j}\sim\epsilon_{{\bf k+q}'',l}:
\nonumber\\
&
-\frac{1}{2}\tilde\delta^{(1)}_{{\bf k},i}.
\end{align}
Finally, in the fifth line of Eq.~\ref{eq_zmet}, when 
$\epsilon_{{\bf k}i} \sim \epsilon_{{\bf k+q}'',j}$ one can substitute
$(\tilde\delta_{{\bf k}i} - \tilde\delta_{{\bf k+q}'',j})
/(\epsilon_{{\bf k}i} - \epsilon_{{\bf k+q}'',j})$
with $-\tilde\delta_{{\bf k},i}^{(1)}$.

Once the the derivative in Eq.~\ref{eqthird} has been
determined, one can obtain the phonon scattering coefficients by combining
Eq.~\ref{eqv3} and Eq.~\ref{eq5}.
Provided that {\bf G} is vector of the reciprocal lattice, we also remark that
\begin{align}
\frac{\partial^3 {\cal E}^{\rm tot}}{\partial u_{{\bf q}}\partial
  u_{{\bf q}'}\partial u_{{\bf q}''+{\bf G}}}=
\frac{\partial^3 {\cal E}^{\rm tot}}{\partial u_{{\bf q}}\partial
  u_{{\bf q}'}\partial u_{{\bf q}''}}
\label{eq_qG}
\end{align}
Hence we have not lost of generality by imposing
${\bf q}+{\bf q}'+{\bf q}''={\bf 0}$ at the beginning of the
present section.

Given a computer code which implements linear response to DFT (DFPT),
all the bra-ket products described in this section can be obtained straightforward.
On the other hand, the computation of the second to the fourth terms
in the r.h.s of Eq.~\ref{eqa12} have to be implemented from scratch.
The implementation of the fourth term is trivial within the local
density approximation. The expressions required for the other terms
are given below.

\subsection{Ionic contribution}
\label{seca4}

The second term in the r.h.s of Eq.~\ref{eqa12} is the third derivative of the ion-ion contribution to
the total energy. It is computed, as customary, using the Ewald sum technique~\cite{ewald}:
\begin{widetext}
\begin{align}
\frac{\partial^3 E^{\rm ion}}{
\partial u_{{\bf q},s,\alpha}
\partial u_{{\bf q}',s',\beta}
\partial u_{{\bf q}'',s'',\gamma}}= {} &
\delta_{s',s''} Z_{s'} Z_{s}
F_{\alpha,\beta,\gamma}({\bf q},{\bf t}_{s'}-{\bf t}_{s}) +
\delta_{s'',s} Z_{s''} Z_{s'}
F_{\alpha,\beta,\gamma}({\bf q}',{\bf t}_{s''}-{\bf t}_{s'})  \nonumber \\
&+\delta_{s,s'}Z_{s} Z_{s''}
F_{\alpha,\beta,\gamma}({\bf q}'',{\bf t}_{s}-{\bf t}_{s''}) -
\delta_{s,s',s''}Z_{s}\sum_{\tilde{s}}Z_{\tilde{s}}
F_{\alpha,\beta,\gamma}({\bf 0},{\bf t}_{s}-{\bf t}_{\tilde{s}}).
\label{eqa16}
\end{align}
\end{widetext}
In Eq.~\ref{eqa16}, we have written explicitly the dependence on the atomic 
($s$,$s'$,$s''$) and Cartesian ($\alpha$,$\beta$,$\gamma$) indexes
of the phonon patterns $u_{{\bf q}}$ (defined in Eq.~\ref{equq}).
In Eq.~\ref{eqa16}, $Z_s$ is the ionic charge and
${\bf t}_s$ is the position of atom $s$. The sum is performed
over all the atoms of the unit cell.
The function $F$ is
\begin{align}
F_{\alpha,\beta,\gamma}({\bf q},{\bf t})= {} &
-\frac{4\pi}{\Omega}
\sum\limits_{{\bf G}}\Bigg[
\frac{e^{-({\bf G+q})^{2}/(4\eta^{2})}}{({\bf G+q})^{2}}
e^{i({\bf G+q})\cdot{\bf t}}
\nonumber \\
&\times
i^3 ({\bf G+q})_\alpha({\bf G+q})_\beta({\bf G+q})_\gamma \Bigg]
\nonumber \\
&-\sum\limits_{{\bf R}}e^{i{\bf q}\cdot{\bf R}}
\left.\frac{d^{3}f({\bf x})}{dx_{\alpha}dx_{\beta}dx_{\gamma}}\right|_{{\bf x=t-R}}.
\end{align}
Here, $\Omega$ is the unit-cell volume,
the sums are performed on the ensemble of the reciprocal lattice vectors {\bf G}
and of the real space lattice vectors {\bf R}.
$\eta$ is the cutoff for the real space summation within the Ewald method~\cite{ewald}
and $f({\bf x})=\mathrm{erfc}(\eta |{\bf x}|)/|{\bf x}|$, being
$\mathrm{erfc}$ is the error function. The derivative of $f$ is
\begin{align}
\frac{d^{3}f({\bf x})}{dx_{\alpha}dx_{\beta}dx_{\gamma}}  = {}&
(\delta_{\alpha\beta}x_{\gamma}+\delta_{\beta\gamma}x_{\alpha}+\delta_{\gamma\alpha}x_{\beta})f_{1}(|{\bf
x}|) \notag\\
   & +x_{\alpha}x_{\beta}x_{\gamma}f_{2}(|{\bf x}|)\,
\end{align}
with
\begin{align}
f_{1}(x) &= \frac{3\mathrm{erfc}(\eta x)+a(\eta x)(3+2\eta^{2}x^2)}{x^{5}}, \nonumber\\
f_{2}(x) &= -\frac{15\mathrm{erfc}(\eta x)+a(\eta x)(15+10\eta^{2}x^{2}+4\eta^{4}x^{4})}{x^{7}}, \nonumber\\
a(\xi) &= \frac{2\xi}{\sqrt{\pi}}e^{-\xi^{2}} \nonumber.
\end{align}

\subsection{Derivatives of the external potential}
\label{seca5}

Here we give the expressions to calculate the 
third and fourth terms in the r.h.s of Eq.~\ref{eqa12}.
Both terms are convenient to evaluate in the reciprocal space.
$v^{\rm ion}$ in Eq.~\ref{eqa12} corresponds to the electrostatic
potential induced by the atomic ions. Within the present approach
an atom $s$ acts through a pseudopotential, which, within the
Kleinman-Bylander scheme ~\cite{kleinman82},
is nonlocal and can be written
\begin{equation}
v_s({\bf r},{\bf r}')=v^{\rm loc}_s({\bf r})\delta({\bf r}-{\bf r}')+
\sum\limits_{\mu,\nu}D^s_{\mu,\nu}P_{\mu,s}({\bf r})P_{\nu,s}({\bf r}'),
\end{equation}
where $v^{\rm loc}_s$ is the local component of the potential;
$D^s_{\mu,\nu}=D^s_{\nu,\mu}$ 
are coefficients and $P_{\mu,s}$ are ion-centered projectors. The total ionic potential
is a superposition of ionic potentials:
\begin{equation}
v^{\rm ion}({\bf r},{\bf r'})=\sum\limits_{{\bf R},s}
v_s({\bf r}-{\bf t}_s-{\bf R},{\bf r'}-{\bf t}_s-{\bf R}),
\end{equation}
where the sum is performed on all the lattice vectors {\bf R}
and on the atoms $s$ in the unit cell.
When $v^{\rm ion}$ is local, its trace with the charge density is
$\int v^{\rm ion}({\bf r})n({\bf r})d{\bf r}$.
When $v^{\rm ion}$ is nonlocal, the same quantity is
$\sum\limits_{{\bf k},i}\tilde\theta_{{\bf k},i}
\int d{\bf r}\int d{\bf r}'
\psi^*_{{\bf k},i}({\bf r}) v^{\rm ion}({\bf r},{\bf r}')\psi^*_{{\bf
    k},i}({\bf r}')$.
With this notation
\begin{align}
\int n({\bf r}) v^{\rm ion}({\bf r}) d{\bf r}={}&
\frac{1}{\Omega}\sum\limits_{{\bf G},s}n(-{\bf G})v^{\rm loc}_s({\bf
  G})e^{-i{\bf G}\cdot{\bf t}_s}\nonumber \\
&+\frac{1}{N}\sum\limits_{\overset{{\bf k},i}{\mu,\nu,s}}
D^s_{\mu,\nu}\tilde\theta_{{\bf k},i}
A_{\overset{{\bf k},i}{\mu,s}}
\Big[A_{\overset{{\bf k},i}{\nu,s}}\Big]^*,
\end{align}
\begin{equation}
A_{\overset{{\bf k},i}{\mu,s}}=
\frac{1}{\Omega}\sum\limits_{\bf G}
\psi^*_{{\bf k},i}({\bf G})P_{\mu,s}({\bf k}+{\bf G})
e^{-i{\bf G}\cdot{\bf t}_s}.
\end{equation}
Here, $n({\bf k})$, $v^{\rm loc}_s({\bf k})$, $P_{\mu,s}({\bf k})$
are the Fourier transform of
$n({\bf r})$, $v^{\rm loc}_s({\bf r})$, $P_{\mu,s}({\bf r})$.
The Fourier transform of $f({\bf r})$ is defined as
$f({\bf k})=\int e^{-i{\bf k}\cdot{\bf r}}f({\bf r}) d{\bf r}$, where
the integral is done all over the space.
Given a Bloch wavefunction $\psi_{{\bf k},i}({\bf r})$, we define
$\psi_{{\bf k},i}({\bf G})=\int e^{-i({\bf k+G})\cdot{\bf r}}\psi_{{\bf
    k},i}({\bf r}) d{\bf r}$, that is
$\psi_{{\bf k},i}({\bf r})=1/(N\Omega)\sum\limits_{\bf G}
e^{i({\bf k+G})\cdot{\bf r}}\psi_{{\bf k},i}({\bf G})$.

To simplify the notation we notice that
the derivative of the charge,
Eq.~\ref{eqlr3}, can be rewritten as
\[
\frac{\partial n}{\partial u_{\bf q}}({\bf r})=
\sum\limits_{{\bf k},i}\tilde\theta_{{\bf k},i}
\Big\{
\tilde\phi^{{\bf q}}_{{\bf k},i}({\bf r})\psi^*_{{\bf k},i}({\bf r})+
\psi_{{\bf k},i}({\bf r})
\big[
\tilde\phi^{-{\bf q}}_{{\bf k},i}({\bf r})
\big]^*\Big\},
\]
and we define
\[
\tilde\phi^{\bf q}_{{\bf k},i}({\bf G})=\int e^{-i({\bf k+q+G})\cdot{\bf
    r}}\tilde\phi^{\bf q}_{{\bf k},i}({\bf r}) d{\bf r}.
\]
By using the above definitions,
the third and the fourth terms in the r.h.s of Eq.~\ref{eqa12} are

\begin{widetext}
\begin{align}
\int n({\bf r})
\frac{\partial^3 v^{\rm ion}({\bf r})}{
\partial u_{{\bf q},s,\alpha}
\partial u_{{\bf q}',s',\beta}
\partial u_{{\bf q}'',s'',\gamma}}d{\bf r} 
={}&
\frac{\delta_{s,s',s''}}{\Omega}\sum\limits_{{\bf G}}n(-{\bf G})
(-i)^3G_\alpha G_\beta G_\gamma v^{\rm loc}_s({\bf
  G})e^{-i{\bf G}\cdot{\bf t}_s}+
\frac{\delta_{s,s',s''}}{N}
\sum\limits_{\overset{{\bf k},i}{\mu,\nu}}
D^s_{\mu,\nu}\tilde\theta_{{\bf k},i}
\nonumber \\
&\times
\Bigg\{
A^{(\alpha,\beta,\gamma)}_{\overset{{\bf k},i}{\mu,s}}
\Big[A_{\overset{{\bf k},i}{\nu,s}}\Big]^*+
\overline{
A^{(\alpha,\beta)}_{\overset{{\bf k},i}{\mu,s}}
\Big[A^{(\gamma)}_{\overset{{\bf k},i}{\nu,s}}\Big]^*}+
\overline{
A^{(\alpha)}_{\overset{{\bf k},i}{\mu,s}}
\Big[A^{(\beta,\gamma)}_{\overset{{\bf k},i}{\nu,s}}\Big]^*}+
A_{\overset{{\bf k},i}{\mu,s}}
\Big[A^{(\alpha,\beta,\gamma)}_{\overset{{\bf k},i}{\nu,s}}\Big]^*
\Bigg\},
\label{eq_nv3}
\end{align}
\begin{align}
\int \frac{\partial n({\bf r})}{\partial u_{{\bf q},s,\alpha}}
\frac{\partial^2 v^{\rm ion}({\bf r})}{
\partial u_{{\bf q}',s',\beta}
\partial u_{{\bf q}'',s'',\gamma}}
d{\bf r} ={}&
\frac{\delta_{s',s''}}{\Omega}\sum\limits_{{\bf G}}
\frac{\partial n(-{\bf G})}{\partial u_{-{\bf q},s,\alpha}}
(-i)^2({\bf q+G})_\beta({\bf q+G})_\gamma
v^{\rm loc}_{s'}({\bf q+G})e^{-i({\bf q+G})\cdot{\bf t}_{s'}}+
2\frac{\delta_{s',s''}}{N}
\sum\limits_{\overset{{\bf k},i}{\mu,\nu}}
\nonumber \\
&\times
D^{s'}_{\mu,\nu}
\tilde\theta_{{\bf k},i}
\Bigg\{
A^{(\beta,\gamma)}_{\overset{{\bf k},i}{\mu,s'}}
\Big[B_{\overset{{\bf q},{\bf k},i}{\nu,s'}}\Big]^*+
A^{(\beta)}_{\overset{{\bf k},i}{\mu,s'}}
\Big[B^{(\gamma)}_{\overset{{\bf q},{\bf k},i}{\nu,s'}}\Big]^*+
A^{(\gamma)}_{\overset{{\bf k},i}{\mu,s'}}
\Big[B^{(\beta)}_{\overset{{\bf q},{\bf k},i}{\nu,s'}}\Big]^*+
A_{\overset{{\bf k},i}{\mu,s'}}
\Big[B^{(\beta,\gamma)}_{\overset{{\bf q},{\bf k},i}{\nu,s'}}\Big]^*
\Bigg\}.
\label{eq_n1v2}
\end{align}
\end{widetext}

In Eq.~\ref{eq_nv3}, the overline
indicates the sum on the permutations of the
Cartesian indexes, \textit{e.g.}
$\overline{A^{\alpha\beta}[A^{\gamma}]^*}=
A^{\alpha\beta}[A^{\gamma}]^*+
A^{\beta\gamma}[A^{\alpha}]^*+
A^{\gamma\alpha}[A^{\beta}]^*$.
Furthermore, 
$\frac{\partial n({\bf G})}{\partial u_{{\bf q}}}$
is the Fourier transform of
$\frac{\partial n({\bf r})}{\partial u_{{\bf q}}}$.
The $A$ and $B$ coefficients of 
Eqs.~\ref{eq_nv3} and ~\ref{eq_n1v2} are defined as

\begin{align}
A^{(\alpha)}_{\overset{{\bf k},i}{\mu,s}}={}&
\frac{-i}{\Omega}\sum\limits_{\bf G}
\psi^*_{{\bf k},i}({\bf G})
e^{-i{\bf G}\cdot{\bf t}_s}
({\bf k +G})_\alpha
P_{\mu,s}({\bf k}+{\bf G}),
\nonumber \\
A^{(\alpha,\beta)}_{\overset{{\bf k},i}{\mu,s}}={}&
\frac{(-i)^2}{\Omega}\sum\limits_{\bf G}
\psi^*_{{\bf k},i}({\bf G})
e^{-i{\bf G}\cdot{\bf t}_s}
\nonumber\\
&\times ({\bf k +G})_\alpha ({\bf k +G})_\beta
P_{\mu,s}({\bf k}+{\bf G}),
\nonumber\\
A^{(\alpha,\beta,\gamma)}_{\overset{{\bf k},i}{\mu,s}}={}&
\frac{(-i)^3}{\Omega}\sum\limits_{\bf G}
\psi^*_{{\bf k},i}({\bf G}) e^{-i{\bf G}\cdot{\bf t}_s}
\nonumber\\
&\times({\bf k +G})_\alpha ({\bf k +G})_\beta ({\bf k +G})_\gamma
P_{\mu,s}({\bf k}+{\bf G}).
\nonumber
\end{align}
\begin{align}
B_{\overset{{\bf q},{\bf k},i}{\mu,s}}={}&
\frac{1}{\Omega}\sum\limits_{\bf G}
\big[\tilde\phi^{\bf q}_{{\bf k},i}({\bf G})\big]^*
P_{\mu,s}({\bf k + \bf q}+{\bf G})
e^{-i({\bf q+G})\cdot{\bf t}_s},
\nonumber\\
B^{(\alpha)}_{\overset{{\bf q},{\bf k},i}{\mu,s}}={}&
\frac{-i}{\Omega}\sum\limits_{\bf G}
\big[\tilde\phi^{\bf q}_{{\bf k},i}({\bf G})\big]^*
({\bf k + q + G})_\alpha
\nonumber\\ & \times 
P_{\mu,s}({\bf k}+{\bf q}+{\bf G})
e^{-i({\bf q+G})\cdot{\bf t}_s},
\nonumber\\
B^{(\alpha,\beta)}_{\overset{{\bf q},{\bf k},i}{\mu,s}}={}&
\frac{(-i)^2}{\Omega}\!\!\sum\limits_{\bf G}
\big[\tilde\phi^{\bf q}_{{\bf k},i}({\bf G})\big]^*
({\bf k +  q + G})_\alpha({\bf k +  q + G})_\beta
\nonumber\\ & \times
P_{\mu,s}({\bf k}+{\bf q}+{\bf G})
e^{-i({\bf q+G})\cdot{\bf t}_s}.
\nonumber
\end{align}
Where the notation $({\bf k+G})_\alpha$ means: The cartesian component $\alpha$ of vector ${\bf k+G}$.

\subsection{Nonlinear core correction}
\label{seca6}

Within the pseudopotential approach, the electronic charge density is
divided into core and valence contributions.
A common way to include the core effects in the Kohn-Sham equations
is the nonlinear core correction scheme of Ref.~\onlinecite{louie82}.
In practice, one determines the charge density of the core electrons
$n_{\rm c}({\bf r})$ only once, before the KS self-consistent cycle, 
as $n_{\rm c}({\bf r})$ only depends on the pseudopotentials.
The KS equations are still solved only for the valence bands and $n({\bf
r})$ defined in Eq.~\ref{eqks3} remains the valence charge.
On the contrary, in Eq.~\ref{eqks2} one substitutes $E_{\rm I}[n]$
with $E_{\rm I}[n_{\rm t}]$ where $n_{\rm t}({\bf r})=n({\bf
  r})+n_{\rm c}({\bf r})$ is the total charge density.

By using this approach the third order equations need to be modified.
First, in the last term in the right hand side of Eq.~\ref{eqa12},
one has to replace the three $\partial n/\partial u_{\bf q}$
with the corresponding $\partial n_{\rm t}/\partial u_{\bf q}$.
Second, one has to include in the r.h.s. of Eq.~\ref{eqa12}
the two additional terms
\begin{equation}
\begin{split}
{}&\int\frac{\delta E_{\rm I}[n_{\rm t}]}{\delta n({\bf r})}
\frac{\partial^3n_{\rm c}({\bf r})}
{\partial u_{{\bf q}}\partial u_{{\bf q}'}\partial u_{{\bf
      q}''}}d{\bf r} \\
{}&+
3\int\frac{\delta^2 E_{\rm I}[n_{\rm t}]}{\delta n({\bf r})\delta n({\bf r'})}
\frac{\partial^2n_{\rm c}({\bf r})}
{\partial u_{{\bf q}}\partial u_{{\bf q}'}}
\frac{\partial^2n_{\rm t}({\bf r'})}{\partial u_{{\bf q}''}} d{\bf r}d{\bf r'}
\end{split}
\end{equation}

\section{Notes on the implementation}
\label{app_implementation}
This section discusses some practical issues concerning the actual
implementation of the method.

\subsection{Implementation}

Let us consider three wavevectors in the general case in which
${\bf q}\ne{\bf q}'\ne{\bf q}''$, with
${\bf q}+{\bf q}'+{\bf q}''={\bf 0}$.
The calculations of the third order derivatives is done in three
consecutive steps.

\begin{enumerate}
\item Self consistent calculation to obtain the
ground state charge $n({\bf r})$ and the Kohn-Sham potential $V^{\rm KS}({\bf r})$.
\item Self consistent linear response
calculation to determine $\partial n ({\bf r})/\partial u_{\tilde {\bf q}}$
and $\partial V^{\rm KS} ({\bf r})/\partial u_{\tilde {\bf q}}$.
This is done three times for the three cases
$\tilde{\bf q}={\bf q}'$, ${\bf q}''$, and ${\bf q}''$.
\item Calculation of the third order derivatives.
\end{enumerate}

In order to do Step 3,
seven distinct sets of ground-state wavefunctions $|\psi\rangle$ are needed (see eq. \ref{eq_ins} or \ref{eq_zmet}):
$|\psi_{\bf k}\rangle$, $|\psi_{\bf k+\tilde q}\rangle$, $|\psi_{\bf k-\tilde q}\rangle$,
with $\tilde{\bf q}$={\bf q}, ${\bf q}'$, and ${\bf q}''$, where {\bf k} runs on the
ensemble of wavevectors which are used in the electronic integration.
The $|\psi_{\bf k}\rangle$ are already calculated in Step 1. They can be saved
and read from disk or they can be calculated a second time with a relatively
inexpensive non self-consistent solution of the KS equation
(based on the knowledge of $V^{\bf KS}$ and $n({\bf r})$ determined in Step 1).
When {\bf k} runs on a regular grid and $\tilde {\bf q}$ belongs
to the grid, the different sets might coincide. However, given the way the code is implemented
~\cite{qe} (the wavefunctions are read and written from sequential files;
different processors might work with different {\bf k} points in parallel)
it is bests to keep distinct the seven sets.

Twelve distinct sets of wavefunctions derivatives  are also needed:
$|\phi^{\tilde q}_{\bf k}\rangle$,
$|\phi^{-\tilde q}_{\bf k}\rangle$,
$|\phi^{\tilde q}_{\bf k-\tilde q}\rangle$, and
$|\phi^{-\tilde q}_{\bf k+\tilde q}\rangle$, 
with $\tilde{\bf q}$={\bf q}, ${\bf q}'$, and ${\bf q}''$.
These wavefunctions can be calculated with a non self-consistent solution of 
the linearized KS equation which is based on the knowledge of the
$\partial V^{\rm KS} ({\bf r})/\partial u_{\tilde {\bf q}}$ calculated in Step 2.
Even in this case, for practical reasons, it is better to keep distinct the 
twelve sets even when some of them coincide.
These derivatives are computed resolving non-self-consistently eq. \ref{eqks1};
in the present work, this was the most CPU intensive step; on the other hand, for bigger systems the 
Input/Output of $\partial n/\partial u_{\bf q}$ can become a bottleneck.

In order to evaluate Eq.~\ref{eq_ins}, one needs to calculate matrix elements
of the kind $\langle\phi^{\bf -q}_{\bf k}|V^{{\bf q}'}|\phi^{{\bf q}''}_{\bf k}\rangle$
and $\langle\psi_{\bf k-q'}|V^{{\bf q}''}|\psi_{\bf k+q}\rangle$.
In both cases this calculation has to be done six times for all the possible
permutations of the three wavevectors ${\bf q}$, ${\bf q}'$, ${\bf q}''$.
For the metallic case, Eq.~\ref{eq_zmet}, further terms are needed. The terms
$\langle\phi^{{\bf -q}_a}_{{\bf k+q}_a}|V^{{\bf q}_b}|\psi_{{\bf k}+{\bf q}_b}\rangle$
need to be computed six times for ${\bf q}_a$ and ${\bf q}_b$ equal
to ${\bf q}$,${\bf q'}$, ${\bf q}''$, with ${\bf q}_a\ne{\bf q}_b$.
The terms $\langle\psi_{\bf k+q}|V^{{\bf q}}|\psi_{\bf k}\rangle$ and
$\langle\psi_{\bf k-q}|V^{{\bf -q}}|\psi_{\bf k}\rangle$ are computed already in Step 2 and
can be saved for later use in Step 3.
The terms $\langle\psi_{\bf k-q}|V^{{\bf q'}}|\psi_{\bf k+q''}\rangle$ need to be computed
for the six permutations of ${\bf q}$, ${\bf q}'$, and ${\bf q}''$.

In some special cases the number of operations can be greatly reduced.
The most obvious one is when ${\bf q}={\bf q}'={\bf q}''={\bf 0}$: in this case
only one set of wavefunctions derivatives is needed: $|\phi^{\bf 0}_{\bf k}\rangle$.
Moreover, when ${\bf q}={\bf 0}$, ${\bf q}'=-{\bf q}''={\bf p}$, with
${\bf p}\ne{\bf 0}$, only three sets of wavefunctions derivatives
are needed: $|\phi^{\bf 0}_{\bf k}\rangle$, $|\phi^{\bf q}_{\bf k}\rangle$, and $|\phi^{\bf q}_{\bf k-q}\rangle$.
This is because $\partial n({\bf r})/\partial u_{\bf -q}=
[\partial n({\bf r})/\partial u_{\bf q}]^*$ and, according to time reversal symmetry,
$\phi^{\bf -q}_{\bf -k}({\bf r})=[\phi^{\bf q}_{\bf k}({\bf r})]^*$.
Another special case is when 
${\bf q}=2{\bf p}$, ${\bf q}'={\bf q}''=-{\bf p}$; in this case 8 derivatives are needed and 
index permutation can be used to avoid computing some terms.

\subsection{Symmetry and q-points}

A Bravais lattice can be invariant under a certain number, up to 48, of rotations.
A crystal will respect a subset, possibly all, of these symmetries
defining the group of crystal symmetries, ${\mathcal G}$.
These symmetries can be exploited to reduce the computational cost of
the calculation and must be enforced to avoid unphysical breaking of
symmetry caused by computational noise.
Here, we briefly revise how this is done in the
{\sc pwscf} and {\sc phonon} codes of {\sc Quantum ESPRESSO}, and
we describe how the approach has been extended to the third order.

In the ground-state energy calculation,
the Kohn-Sham equations (Eqs. ~\ref{eqks1}, ~\ref{eqks2}, ~\ref{eqks3}),
are not solved for all the {\bf k}-points of the chosen grid. Instead
they are solved only for a subset of {\bf k}-points which are
inequivalent under the rotations of ${\mathcal G}$
(these {\bf k} points belong to the so called irreducible wedge of the Brillouin zone).
The charge density $n({\bf r})$ is then obtained, inexpensively, by summing on this
{\bf k}-point subset and, then, by imposing the symmetries of the
crystal. The symmetries are imposed by rotating the initial charge density
according to all the symmetries and, then, by making the average.
In general, even if the crystal has no symmetries, the sum can be performed on
half of the {\bf k}-points of the initial grid since
$\psi_{\bf -k}({\bf r}) = \psi^*_{\bf k}({\bf  r})$ (time reversal symmetry).

In the phonon calculation, the situation depends on the phonon wave-vector {\bf q}.
Let us consider the the symmetries of the crystal which are associated
with a rotation that leaves {\bf q} unchanged
(or that transform ${\bf q}$ into an equivalent ${\bf q+G}$).
This subgroup of ${\mathcal G}$ is called the small group of {\bf q}, ${\mathcal G}_{\bf q}$.
The linear response equations (Eqs. ~\ref{eqlr1}, ~\ref{eqlr2}, ~\ref{eqlr3})
are not solved on the grid. Instead, they are solved only
for a subset of {\bf k}-points which are inequivalent under the rotations
of ${\mathcal G}_{\bf q}$.
$\partial n({\bf r})/\partial u_{\bf q}$ is obtained by summing on this
{\bf k}-point subset and, then, by imposing the symmetries of
${\mathcal G}_{\bf q}$ (in analogy with $n({\bf r})$ in the previous paragraph).
Because of time reversal symmetry, in order to
obtain $\partial n({\bf r})/\partial u_{\bf q}$ from Eq.~\ref{eqlr3}
one needs to compute explicitly only the $|\phi^{\bf q}_{\bf k}\rangle$ but not the
$|\phi^{\bf- q}_{\bf k}\rangle$. However, if there are no other
symmetries, the sum has to be performed on all the {\bf k}-points of
the grid and not on half as in the total energy calculation.
Once that $\partial n({\bf r})/\partial u_{\bf q}$ has been calculated, one can
determine the dynamical matrix at {\bf q}.

Another way in which the symmetries can be exploited is the following.
The ensemble of the vectors obtained
by rotating {\bf q} with all the rotations of ${\mathcal G}$
is called the star of {\bf q}.
The number of inequivalent vectors in the star is 
$|{\mathcal G}|/|{\mathcal G}_{\bf q}|$, where $|{\mathcal G}|$ is the
order of ${\mathcal G}$.
Once that $\partial n({\bf r})/\partial u_{\bf q}$ has been determined,
all the $\partial n({\bf r})/\partial u_{\tilde{\bf q}}$ for the
vectors  $\tilde{\bf q}$ in the star can be obtained inexpensively by rotation.
Moreover, because of time reversal symmetry
$\partial n({\bf r})/\partial u_{-{\bf q}}=
[\partial n({\bf r})/\partial u_{{\bf q}}]^*$.

A third-order calculation is conceptually not different.
In this case we are dealing with a triplet of points
$({\textbf q},{\textbf q}',{\textbf q}'')$ and one has to consider
the small group of the rotations of ${\mathcal G}$ that
leave the three {\bf q}-points unchanged, 
${\mathcal G}_{{\bf q,q',q''}}$.
The electronic {\bf k}-point summation in Eq.~\ref{eq_ins}, ~\ref{eq_zmet}
(and also those in Eqs.~\ref{eq_nv3}, ~\ref{eq_n1v2})
is done on a subset of point of the initial grid
which are inequivalent under the rotations of ${\mathcal G}_{{\bf q,q',q''}}$.
The third order matrix obtained from this partial summation is not
useful as it is. The actual third order matrix is obtained by imposing
on it the symmetries of ${\mathcal G}_{{\bf q,q',q''}}$.

In analogy to the dynamical matrix case, once the third-order
matrix of a given $({\textbf q},{\textbf q}',{\textbf q}'')$ triplet has been
determined, on can apply the rotations ${\mathcal R}$ of the crystal symmetry
group ${\mathcal G}$ and obtain inexpensively 
the matrices corresponding to the rotated triplet 
$({\mathcal R}{\bf q},{\mathcal R}{\bf q'},{\mathcal R}{\bf q''})$.
Moreover, the matrix of the triplet
$(-{\textbf q},-{\textbf q}',-{\textbf q}'')$ can be obtained by
conjugation.
Another useful symmetry is the trivial one associated to the index
permutation: by construction, the matrix associated with the triplet
$({\textbf q},{\textbf q}',{\textbf q}'')$
is equal to the six matrices obtained by permuting the three
vectors ${\bf q}$,${\bf q}'$,${\bf q}''$ and the corresponding indices.

\subsection{Discretisation on a grid}
\label{app_grid}

In some cases, it is useful to calculate the third-order matrices on a
regular grid of {\bf q} wavevectors in the Brillouin zone,
the reason for this are clarified in Sec.~\ref{app2}.
When we say that calculations are done on a given grid, it means
that we have calculated the third order coefficients corresponding to
every triplet $({\bf q},{\bf q}',{\bf q}'')$ of points, such that 
each vector belongs to the grid and that the condition 
${\bf q}+{\bf q}'+{\bf q}''=G$ is satisfied.

In the actual implementation of the procedure, we first determine
$\partial n({\bf r})/\partial u_{\bf q}$ and the dynamical matrices
on the grid of {\bf q} wavevector.
Exploiting the symmetries can allow an important reduction of the
computational cost. Indeed,
the actual ab-initio calculation is done only for those
{\bf q} points which do not belong to the same star (defined in the previous section),
or which are not equivalent to the opposite of a point already
calculated.
$\partial n({\bf r})/\partial u_{\bf q}$ and the dynamical matrices
for the other points are then obtained by rotation or conjugation.
Once this is done we perform the third order calculations.

In practice, we can perform a double loop on the grid and
chose ${\bf q}$ and ${\bf q}'$ so that they belong to the grid.
The third vector is chosen as ${\bf q}'' =-{\bf q}-{\bf q}'$.
${\bf q}''$ may not belong to the grid. ${\bf q}''$ is however still connected
to a point in grid by a reciprocal lattice vector {\bf G} and, thanks
to Eq. ~\ref{eq_qG}, we are not losing generality.
Moreover,
\[
\frac{\partial n({\bf r})}{\partial u_{\bf q''}}=\frac{\partial n({\bf r})}{\partial u_{\bf q''+G}}
\]
for every reciprocal space vector {\bf G}, and we can use the 
$\partial n({\bf r})/\partial u_{\bf q}$ calculated on the original grid.
Once that a triplet has been computed we can determine for free all the triplets
which are equivalent by rotation, permutation of the indices, or by conjugation.
Actually, some of these operations can be redundant,
\textit{e.g.} a certain rotation could be equivalent to a permutation, or to
a conjugation.
The triplet obtained in this way are deleted from the list of the
``triplets to be done'' and will not be computed in the following step
of the loop. This procedure allows for a spectacular reduction in the number of
triplets; in the graphene case 4096 possible triplets are reduced
to just 88 independent triplets.

\section{Fourier interpolation}
\label{app2}
Actual DFPT calculations are done on a relatively coarse
grid of {\bf q} wavevectors.
Dynamical matrices and third order
coefficients are then obtained for a finer grid with a Fourier interpolation technique.
In this section, first, we revise the Fourier interpolation technique as it is 
implemented in the standard {\sc Quantum ESPRESSO} package.
Then, we describe how the method has been generalized to third order
force constants.

\subsection{Second order}
Let us consider a lattice with basis ${\bf a}_1$, ${\bf a}_2$, ${\bf a}_3$.
The dynamical matrices 
$D_2\left(
\begin{smallmatrix}
&{\bf q}\\
s&s'
\end{smallmatrix}
\right)$ 
(we use the definition of Eq.~\ref{eqdyn} and we drop for simplicity
the Cartesian indexes) are first computed {\it ab initio}
on a uniform grid, centered in the origin, of $N_1\times N_2\times N_3$ {\bf q} points of the
Brillouin zone.
We want to determine the real space force constants
\begin{equation}
F_2\left(
\begin{smallmatrix}
&{\bf R}\\
s&s'
\end{smallmatrix}
\right)
=\frac{\partial^2 {\cal E}^{\rm tot}}{\partial v_{{\bf 0},s}\partial
  v_{{\bf R},s'}}
\label{eqb1}
\end{equation}
by Fourier interpolation of the $D_2$ from the $N_1\times N_2\times N_3$ grid. 
$D_2$ for a generic {\bf q} point can then be obtained
by back Fourier interpolation.

Let us consider
\begin{equation}
\tilde F_2\left(
\begin{smallmatrix}
&{\bf R}\\
s&s'
\end{smallmatrix}
\right)
=\frac{1}{N_{\rm t}} \sum_{\bf q}
D_2\left(
\begin{smallmatrix}
&{\bf q}\\
s&s'
\end{smallmatrix}
\right)e^{-i{\bf q}\cdot{\bf R}},
\end{equation}
where {\bf R} is a lattice vector, $N_{\rm t}=N_1N_2N_3$, and the sum
is performed on the points of the grid.
The $\tilde F_2$ defined in this way cannot be used as they are since
they are unphysical long-ranged constants.
Indeed,
$\tilde F_2\left(
\begin{smallmatrix}
&{\bf R}+\overline{\bf R}\\
s&s'
\end{smallmatrix}
\right)=
\tilde F_2\left(
\begin{smallmatrix}
&{\bf R}\\
s&s'
\end{smallmatrix}
\right)$
for any $\overline{\bf R}$ vector of the super lattice (SL)
generated by the lattice vectors
${\bf A}_1=N_1{\bf a}_1$, 
${\bf A}_2=N_2{\bf a}_2$, 
${\bf A}_3=N_3{\bf a}_2$.

The short-ranged $F_2$ can be obtained from the $\tilde F_2$
in the following way.
We define two vectors as ``SL equivalent'' when their
difference is a vector of the super lattice defined by
${\bf A}_1$, ${\bf A}_2$, ${\bf A}_3$ and we call
$W_{\rm SL}$ as the Wigner-Seitz of this super lattice.
Eq.~\ref{eqb1} is the force constant between the atoms
whose distance is
${\bf d}={\bf R}+{\bf t}_{s'}-{\bf t}_s$, where
${\bf t}_s$ is the position of the atom $s$ in the unit cell.
We distinguish three cases:
i) when ${\bf d}\in W_{\rm SL}$,
$F_2\left( \begin{smallmatrix} &{\bf R}\\ s&s' \end{smallmatrix} \right)=
\tilde F_2\left( \begin{smallmatrix} &{\bf R}\\ s&s' \end{smallmatrix}
\right)$; 
ii) when {\bf d} lies on the border of $W_{\rm SL}$,
$F_2\left( \begin{smallmatrix} &{\bf R}\\ s&s' \end{smallmatrix} \right)=
\tilde F_2\left( \begin{smallmatrix} &{\bf R}\\ s&s' \end{smallmatrix}
\right)/N_{\rm eq}$, where $N_{\rm eq}$ is the number of points on the border of $W_{\rm SL}$
which are ``SL equivalent'' to {\bf d};
iii) when ${\bf d}\notin W_{\rm SL}$,
$F_2\left(
\begin{smallmatrix}
&{\bf R}\\
s&s'
\end{smallmatrix}
\right)=0$.
The interpolated dynamical matrix for a generic {\bf q} is then
\begin{equation}
D_2\left(
\begin{smallmatrix}
&{\bf q}\\
s&s'
\end{smallmatrix}
\right)
=\sum_{\bf R}
F_2\left(
\begin{smallmatrix}
&{\bf R}\\
s&s'
\end{smallmatrix}
\right)e^{i{\bf q}\cdot{\bf R}},
\end{equation}
where the sum is done on all the lattice vectors {\bf R}.
This is the Fourier interpolation technique as it is 
implemented in the standard {\sc Quantum ESPRESSO} package.

\subsection{Third order}
For the third order coefficients the situation is analogous, although
less straightforward.
We use the notation:
\begin{equation}
\begin{split}
D_3\left(
\begin{smallmatrix}
{\bf q}&{\bf q'}&{\bf q''}\\
s&s'&s''
\end{smallmatrix}
\right)
={}&\frac{1}{N}
\frac{\partial^3 {\cal E}^{\rm tot}}
{\partial u_{{\bf q},s}\partial u_{{\bf q'},s'}\partial u_{{\bf q''},s''}},
\\
F_3\left(
\begin{smallmatrix}
{\bf R}&{\bf R'}&{\bf R''}\\
s&s'&s''
\end{smallmatrix}
\right)
={}&\frac{\partial^3 {\cal E}^{\rm tot}}
{\partial v_{{\bf R},s}\partial v_{{\bf R'},s'}\partial v_{{\bf
      R''},s''}}.
\label{eqb4}
\end{split}
\end{equation}

The matrices $D_3$ are first computed {\it ab initio}
on a uniform grid of {\bf q} points centered in the origin, meaning
that both vectors {\bf q}$'$ and {\bf q}$''$ run on the grid, while
${\bf q}=-{\bf q'}-{\bf q''}$.
We then define
\begin{equation}
\tilde F_3\left(
\begin{smallmatrix}
{\bf 0}&{\bf R'}&{\bf R''}\\
s&s'&s''
\end{smallmatrix}
\right)
=\frac{1}{N_{\rm t}^2}\sum_{{\bf q'},{\bf q''}}
D_3\left(
\begin{smallmatrix}
{\bf q}&{\bf q'}&{\bf q''}\\
s&s'&s''
\end{smallmatrix}
\right)e^{-i({\bf q'}\cdot{\bf R'}+{\bf q''}\cdot{\bf R''})},
\end{equation}
where the sums are performed on the grid points.

The force constants $F_3$ of Eq.~\ref{eqb4}
correspond to three atoms at the positions
${\bf t}_s$,
${\bf t}_{s'}+{\bf R'}$, and
${\bf t}_{s''}+{\bf R''}$
(because of the crystal translational symmetry we can consider
without loss of generality {\bf R}={\bf 0}).
The three atoms form a triangle with perimeter
$P=|{\bf d}_1|+|{\bf d}_2|+|{\bf d}_3|$, where
${\bf d}_1$, ${\bf d}_2$, ${\bf d}_3$ are the distances among the
three atoms (the three sides of the triangle).
To determine $F_3$ we consider three cases.
i) When all the three distances ${\bf d}_1,{\bf d}_2,{\bf d}_3$
are inside (and not on the border) of $W_{\rm SL}$, we put
$F_3\left(
\begin{smallmatrix}
{\bf 0}&{\bf R'}&{\bf R''}\\
s&s'&s''
\end{smallmatrix}
\right)=
\tilde F_3\left(
\begin{smallmatrix}
{\bf 0}&{\bf R'}&{\bf R''}\\
s&s'&s''
\end{smallmatrix}
\right)$.
ii) If two of the distances ({\it e.g.} ${\bf d}_1,{\bf d}_2$)
are inside $W_{\rm SL}$ and the third one is outside or on the border of $W_{\rm SL}$
we calculate the perimeter $P$ of the triangle.
If $P$ is the shortest perimeter among the perimeters of all the triangles formed by the 
triplet of atoms ``SL equivalent'' to the original three, we put
$F_3\left(
\begin{smallmatrix}
{\bf 0}&{\bf R'}&{\bf R''}\\
s&s'&s''
\end{smallmatrix}
\right)=
\tilde F_3\left(
\begin{smallmatrix}
{\bf 0}&{\bf R'}&{\bf R''}\\
s&s'&s''
\end{smallmatrix}
\right)/N_{\rm eq}$,
where $N_{\rm eq}$ is the number of triangles with
perimeter equal to $P$.
iii) In all the other cases 
$F_3\left(
\begin{smallmatrix}
{\bf 0}&{\bf R'}&{\bf R''}\\
s&s'&s''
\end{smallmatrix}
\right)=0$.

The interpolated $D_3$ matrices for a generic triplet of wavevectors
(such that ${\bf q}+{\bf q'}+{\bf q''}={\bf G}$) is obtained from
\begin{equation}
D_3\left(
\begin{smallmatrix}
{\bf q}&{\bf q'}&{\bf q''}\\
s&s'&s''
\end{smallmatrix}
\right)=
\sum_{{\bf R'},{\bf R''}}
F_3\left(
\begin{smallmatrix}
{\bf 0}&{\bf R'}&{\bf R''}\\
s&s'&s''
\end{smallmatrix}
\right)e^{i({\bf q'}\cdot{\bf R'}+{\bf q''}\cdot{\bf R''})},
\end{equation}
where the sums run on all the lattice vectors.
The criterion presently described to determine the $F_3$s,
stems from the assumption of a long range exponential decay of the
force constants.

\section{Computational details}
\label{comp_det}

\subsection{Electronic integration}

The electronic integration of the density functional theory calculations
is done using a first-order Methfessel-Paxton
smearing\cite{methfessel89} of $0.02$~Ry which converges for a grid of
$32\times 32\times 1$ electronic {\bf k}-points in simple and bilayer
graphene and for a grid of $32\times 32 \times 8$ {\bf k}-points for
graphite.
\break
\subsection{Linear response calculations}

DFT dynamical matrices are corrected using a procedure based on DFT+GW
renormalization of the electron-phonon interaction as in Ref.~\onlinecite{lazzeri08}.
Indeed, DFT reproduces very well the measured dispersions of graphite
for all the phonon branches but for the TO one,
in the vicinity of the high symmetry point {\bf K}. This failure of DFT, which is very specific to
the graphene and graphite systems, has been analyzed in
Ref.~\onlinecite{lazzeri08}.
To improve the accuracy of the TO phonon branch, we have applied an
electron-phonon self-interaction as described in
Ref.~\onlinecite{lazzeri08}. The detailed procedure is described in
Sec. IIB of Ref.~\onlinecite{venezuela} (third paragraph). We have used
the parameter $r^{GW}=1.65$, which is appropriate to the present
LDA calculations and which can be derived from Table I of Ref.~\onlinecite{lazzeri08}.
Using this approach we determined the dynamical matrices of graphene
on a super-sampled $48\times 48 \times 1$ {\bf q}-point grid.
The matrices for the graphite and for the bilayer are then obtained 
by using as in-plane force constants those of graphene and
as out-of-plane force constants those coming from independent
DFPT calculations on the two systems.

The third order coefficients are obtained in the standard way,
that is without including this self-interaction correction of Ref.~\onlinecite{lazzeri08}.
For graphene, the third-order coefficients are calculated on a $8 \times 8
\times 1$ {\bf q}-point grid (see Sec.~\ref{app_grid}),
which consists in 88 irreducible triplets.
For bulk graphite, third-order coefficients are calculated on a $8 \times 8 \times 2$
grid, which consists in 297 irreducible {\bf q} triplets.
When computing the linewidth, we have tested convergence starting from
the $8\times8\times2$ grid coefficients, finding that the use of the those from the
$4\times 4\times 2$ subset grid (33 inequivalent triplets) does not worsen accuracy.
For bilayer graphene, the third-order coefficients are calculated on a $4 \times 4
\times 1$ {\bf q}-point grid  (12 irreducible triplets).

\subsection{Broadening calculations}

Eqs.~\ref{eq_tauanh} and~\ref{eq_k} are evaluated by performing the sum over a discrete
uniform grid of ${\bf q}$ points randomly shifted from the origin.
The $\delta(x)$ distribution is substituted with the Gaussian function 
$\tilde\delta(x) = e^{-( x/\chi )^2}/(\chi\sqrt{\pi})$, where
$\chi$ is an artificial smearing, independent from {\bf q}.
The results of Sects.~\ref{sec31} and ~\ref{sec32} are obtained by using:
for graphene, a
$1800\times 1800 \times 1$ grid and $\chi=1$~cm$^{-1}$;
for graphite, $600 \times 600 \times 15$ grid and $\chi=5$~cm$^{-1}$;
for the bilayer, a $1200 \times 1200 \times 1$ grid and $\chi=2$~cm$^{-1}$.
The results of Sec.~\ref{sec33} are obtained using: for graphene,
a $128\times128\times1$ grid and $\chi=10$~cm$^{-1}$; for graphite and bilayer,
a $64\times64\times4$ grid and $\chi=10$~cm$^{-1}$.
For each system, the same grid is used to determine the broadening from
Eq.~\ref{eq_tauanh} and the thermal conductivity from Eq.~\ref{eq_k}.
The convergence, has been tested using smaller smearing values and
finer grids at selected temperatures.

\bibliography{paper}

\end{document}